\documentclass[letterpaper,twocolumn,10pt]{article}
\usepackage{usenix2019}
\newif\iffull
\fullfalse

\usepackage{graphicx}
\usepackage{epstopdf}
\usepackage{mathtools}
\usepackage{cleveref}
\usepackage{listings}
\usepackage{array}
\usepackage{breqn}
\usepackage{tikz}
\usepackage{algorithmicx}
\usepackage[Algorithm,ruled]{algorithm}
\usepackage[noend]{algpseudocode}
\usepackage{xspace}
\newcommand{\name}{{\sc D2R}\xspace}

\newcolumntype{P}[1]{>{\centering\arraybackslash}p{#1}}

\usepackage[font=small]{caption}
\usepackage{enumitem}
\usepackage{epsfig}
\usepackage{verbatim}
\usepackage{color}
\usepackage{verbatim}
\usepackage{fancyvrb}
\usepackage{url}

\usepackage{colortbl}
\usepackage{enumitem}
\usepackage[export]{adjustbox}
\usepackage{subfig}
\usepackage{mathtools}
\usepackage{amssymb}
\usepackage{pifont}
\usepackage{amsthm}
\usepackage{multirow}
\usepackage{multicol}
\usepackage{hhline}
\usepackage{wrapfig}
\usepackage{textcomp}
\usepackage[titletoc]{appendix}

\newcommand{\minisection}[1]{\noindent{\bf #1.}\hspace{1mm}}
\newcommand{\secref}[1]{{\S\ref{#1}}}

\newcommand{\compactcaption}[1]{\vspace{-0.5em}\caption{#1}\vspace{-0.5em}}

\newenvironment{compactitemize}
{
   \begin{itemize}[leftmargin=1.5em]
   \vspace{-1ex}
   \setlength{\topsep}{0pt}
   \setlength{\itemsep}{0em}
   \setlength{\parskip}{0pt}
   \setlength{\parsep}{0pt}
}
{
   \vspace{-1ex}
   \end{itemize}
}

\newenvironment{compactenumerate}
{
   \begin{enumerate}[leftmargin=1.5em]
   \vspace{-1ex}
   \setlength{\topsep}{0pt}
   \setlength{\itemsep}{0em}
   \setlength{\parskip}{0pt}
   \setlength{\parsep}{0pt}
}
{
   \vspace{-1ex}
   \end{enumerate}
}

\algtext*{EndWhile}
\algtext*{EndIf}
\algtext*{EndProcedure}
\algtext*{EndFor}
\let\oldReturn\Return
\renewcommand{\Return}{\State\oldReturn}
\makeatletter
\lst@InstallKeywords k{attributes}{attributestyle}\slshape{attributestyle}{}ld
\makeatother

\lstdefinestyle{customc}{
  belowcaptionskip=1\baselineskip,
  breaklines=true,
  xleftmargin=\parindent,
  language=C,
  showstringspaces=false,
  basicstyle=\small\ttfamily,
  keywordstyle=\small\bfseries\ttfamily\color{NavyBlue},
  commentstyle=\itshape\color{black},
  identifierstyle=\ttfamily\color{black},
  stringstyle=\itshape\color{NavyBlue},
  keywords={ map, flatMap, reduce, return, define, gaussian, gauss, step, elif, then, skip, if, else, reduceByKey, sensitiveAttribute, fairnessTarget, def},
moreattributes={ let, where}, 
attributestyle = \itshape\ttfamily\color{attributecolor}
}

\setlist[itemize]{
    topsep=.5ex,
    itemsep=0ex,
    leftmargin=1em,
}

\setlist[description]{
    labelindent=.4cm,
    style=unboxed,
    leftmargin=.4cm,
    font=\itshape,
    topsep=.5ex,
    itemsep=0ex
}

\let\originalleft\left
\let\originalright\right
\renewcommand{\left}{\mathopen{}\mathclose\bgroup\originalleft}
\renewcommand{\right}{\aftergroup\egroup\originalright}

\errorcontextlines\maxdimen

\makeatletter
\newcommand*{\algrule}[1][\algorithmicindent]{\makebox[#1][l]{\hspace*{.5em}\vrule height .75\baselineskip depth .25\baselineskip}}%

\newcount\ALG@printindent@tempcnta
\def\ALG@printindent{%
    \ifnum \theALG@nested>0
        \ifx\ALG@text\ALG@x@notext
            \addvspace{-3pt}
        \else
            \unskip
            \ALG@printindent@tempcnta=1
            \loop
                \algrule[\csname ALG@ind@\the\ALG@printindent@tempcnta\endcsname]%
                \advance \ALG@printindent@tempcnta 1
            \ifnum \ALG@printindent@tempcnta<\numexpr\theALG@nested+1\relax
            \repeat
        \fi
    \fi
    }%
\usepackage{etoolbox}
\patchcmd{\ALG@doentity}{\noindent\hskip\ALG@tlm}{\ALG@printindent}{}{\errmessage{failed to patch}}
\makeatother

\definecolor{redcircle}{RGB}{240,0,0}

\definecolor{purplecircle}{RGB}{128,0,128}

\definecolor{greencircle}{RGB}{20,240,20}

\newcommand*\redc[1]{\tikz[baseline=(char.base)]{%
\node[shape=circle,fill=redcircle,inner sep=1pt] (char) {\color{white}#1};}}

\newcommand*\purplec[1]{\tikz[baseline=(char.base)]{%
\node[shape=circle,fill=purplecircle,inner sep=1pt] (char) {\color{white}#1};}}

\newcommand*\greenc[1]{\tikz[baseline=(char.base)]{%
\node[shape=circle,fill=greencircle,inner sep=1pt] (char) {\color{white}#1};}}

\usepackage[medium, compact]{titlesec}
\titlespacing*{\section}{1pt}{3.5pt}{2pt}
\titlespacing*{\subsection}{1pt}{3pt}{1.5pt}
\titlespacing*{\subsubsection}{1pt}{3pt}{1.5pt}
\crefname{Section}{§}{§§}
\Crefname{Section}{§}{§§}

\newcolumntype{C}[1]{>{\centering\let\newline\\\arraybackslash\hspace{0pt}}m{#1}}

\usepackage{color}

\usepackage{float}

\begin{document}

\setlength{\abovedisplayskip}{5pt}
\setlength{\belowdisplayskip}{5pt}

\date{}

\title{ \name: Dataplane-Only Policy-Compliant Routing Under Failures}
\author{
{\rm Kausik Subramanian, Anubhavnidhi Abhashkumar, Loris D'Antoni, Aditya Akella}\\
University of Wisconsin-Madison
} 

\maketitle


{\bf {\em Abstract--}} In networks today, the data plane handles
forwarding---sending a packet to the next device in the path---and the
control plane handles routing---deciding the path of the packet in the
network. This architecture has limitations. First, when link failures
occur, the data plane has to wait for the control plane to install new
routes, and packet losses can occur due to delayed routing convergence
or central controller latencies. Second, policy-compliance is not
guaranteed without sophisticated configuration synthesis or controller
intervention. In this paper, we take advantage of the recent advances
in fast programmable switches to perform policy-compliant route
computations entirely in the data plane, thus providing fast reactions to
failures. \name, our new network architecture, can provide the
illusion of a network fabric that is always available and
policy-compliant, even under failures. We implement our data plane in
P4 and demonstrate its viability in real world topologies.

\section{Introduction}
With a plethora of performance-sensitive distributed applications
running on datacenter and wide-area networks, the requirements on the
underlying network fabric have become extremely
stringent~\cite{timegone}. In particular, because the fabric
interconnects applications' end-points, there is a push toward making
it highly available.

A key factor that impacts fabric availability from the perspective of
applications is failures. Even in the most well-managed networks, link/switch
failures are common~\cite{datacenterfailures}. A variety of factors
ranging from device crashes/reboots, cabling, buggy hardware/firmware, power
supply issues, etc., can conspire to constantly induce link/switch failures.

The fabric's behavior under failures critically determines its
perceived availability. When a failure occurs in 
today's data center and wide-area network
fabrics, network forwarding attempts to reconverge to
re-establish paths. When the network is still in an unconverged state,
traffic destined to certain endpoints will have no valid route and
will be dropped. This leads to a precipitous performance degradation
for critical applications. Unfortunately, networks can remain in
unconverged states for unreasonable amounts of time; this is true even
if state-of-the-art approaches to route around failures are
employed~\cite{frr,ddc}.

Our primary goal is to design a fabric that provides the illusion of
being {\em always available}. We define this as: \textit{if, under a
  failure scenario, there exist active paths between a
  source-destination pair, then the fabric must route packets through
  some such path without inducing any drops}.

The other major consideration for networks is policy compliance. For
example, Network Function Virtualization (NFV)~\cite{opennf, e2}, a
popular use-case, allows tenants and operators to specify middlebox
chains that traffic between a set of endpoints must traverse for
security and performance considerations. Because a non-trivial
fraction of such middleboxes are now part of network
fabrics~\cite{andromeda,azurefpga}, the network is also tasked with
ensuring correct middlebox traversal. As another example, operators
may desire to employ various network load-balancing schemes---e.g.,
WCMP~\cite{wcmp}---so that the network can effectively spread load
across multiple available paths to avoid queuing and congestion
drops. 

While operators can use various frameworks for policy
compliance~\cite{simple, flowtags, merlin, propane, genesis, zeppelin,
  netkat}, ensuring that {\em policies always hold}, especially when
reacting to failures, is something that no state-of-art
approach achieves. This is our second goal.

We observe that the main obstacle in realizing an ``always available,
policy-compliant'' network is that \emph{recomputing new
  policy-compliant routes under failures is unreasonably
  slow}. Traditionally, recomputation is performed by a centralized or
distributed control plane; in both cases, the computation is off the
fast path of packet forwarding, and therefore slow. The data plane, which lies on the fast
path, is only equipped to perform forwarding based on the control
plane route computations.  

Thus, to meet our goals, we argue for a refactoring of
responsibilities across control and data planes. Specifically, we
argue for performing all route computation entirely in the data plane
for the fastest reaction. Our paper shows that, with technology available
today, it is possible to realize such data plane-only routing that can
instantaneously react to failures in a policy-compliant manner.

Our network architecture, \name\footnote{Pronounced ``detour''.},
leverages recent programmable data planes to this end. Given a view of
the network topology and current state of the links, \name implements
graph traversal algorithms---e.g., Breadth-first Search and
Iterative-Deepening Depth-first Search---completely in the data plane;
our implementation can compute paths to any destination at \emph{near
  line rates}. To propagate failure information for computing active
routes without imposing reconvergence issues, we use the Failure
Carrying Packets protocol~\cite{fcp} to tag each packet header with
the failures it has encountered along its route. \name switches use
the failure information to guide the graph traversal algorithms and
compute active policy-compliant routes.

Because programmable switches today have limited processing stages,
they may not be able to compute the route to the destination in one
pass through the switch. We address this limitation using the
recirculation capabilities of modern switches that allow packets to
be fed back to the switch for additional processing.  Using a
state-of-art hardware switch, we show we can achieve minimal latency
and throughput degradation as long as we impose a low number of
recirculations. Thus, we propose a hierarchical dataplane 
routing scheme that \emph{nearly eliminates}
recirculations and \emph{runs at line rate}.




\minisection{Contributions}
\begin{compactitemize}
    \item \name, a new network architecture that can provide the
      illusion of a fabric that is always available and
      policy-compliant even under failures, by performing routing in
      the data plane (\secref{sec:architecture}).
    \item A P4 implementation of Breadth-first Search and Iterative-Deepening
    Depth-first Search that can run on software and hardware switches,
    coupled with an implementation of the Failure Carrying Packets protocol for
    routing under failures (\secref{sec:dataplanerouting}).
    \item A hierarchical routing scheme in the data plane, which decreases the 
    processing requirements on each switch by splitting route computation across 
    switches (\secref{sec:hierarchy}).
    \item An implementation of the data plane augmentations necessary
    to compute policy compliant paths (\secref{sec:policy}).
    \item An evaluation of \name's end-to-end routing scheme for different topologies 
    and failure scenarios on software and hardware platforms (\secref{sec:evaluation}).
\end{compactitemize}


\section{Routing under Failures}
\label{sec:challanges}

In this section, we present the
challenges faced today w.r.t providing guarantees of connectivity
without packet loss  while complying
with high-level policies. We examine the role of the control plane and
data plane in both these architectures and argue that a refactoring of
roles is needed to address the challenges.

\subsection{Delay in Reconvergence}

A key goal of our work is to ensure that even when multiple failures
occur packets are delivered without drops as long as network paths
exist to packets' destinations. We examine if and how this is possible
today.

{\bf Distributed control planes fall short.} Many networks use
distributed routing protocols that rely on routers exchanging protocol
messages to convey changes in the network topology, for instance, when
link failures occur.  Each router uses these messages to recompute new
forwarding tables to react to its perceived new state of the network.
Until the information about failures propagates to all routers in the
network, and the network has become quiescent, forwarding tables may
not be consistent across routers. During this \emph{convergence}
period---which can last very long~\cite{blink}---severe packet
losses occur when routes become unavailable~\cite{convergenceloss}.


Furthermore, information about failures is passed via advertisements that are
generated and processed by router software control planes. Francois et.
al~\cite{igpconvergence} study the behavior of IS-IS protocol convergence times
based on different parameters: failure detection, link-state packet (LSP)
generation to notify routers of failures, the overhead of flooding LSPs and
processing advertisements at each router's control plane, and updating the RIB
and FIB for each LSP. For a 21 node topology geo-distributed in Europe and USA,
they observe high convergence times of over \emph{200-1000ms} depending on
different control plane parameters---for instance, how the control plane updates
the FIB can vastly change convergence times. In other words, switch/router
control plane software design can further delay convergence, leading to higher
loss rates.

The state-of-the-art approaches to reduce the impact of convergence are (1)
designing loop-free convergence protocols~\cite{loopfree1, loopfree2,
loopfree3}, and (2) using pre-computed backup paths to route around failures,
e.g., LFA-FRR~\cite{frr}, DDC~\cite{ddc} etc. The former approach uses provably
correct mechanisms to break loops during convergence, but can only provide
guarantees for a subset of network failures; it still incur convergence delays
due to the switch software control plane processing. Local fast failover
mechanisms in the data plane---i.e., the latter approach---are widely deployed,
but they only provide guarantees for certain failure scenarios (generally one
link failures); pre-computing backup paths for multiple failure scenarios will
lead to an exponential increase in switch memory usage, or cannot avoid
convergence problems when the failure scenario is not protected by the backup
mechanisms. We performed an empirical analysis of using LFA-FRR (Loop-Free
Alternate Fast Reroute) for protection of every 1-link failure, and we observe
that 2-20\% traffic classes are disconnected until convergence under 2 and 3
link failure scenarios.

{\bf Modern SDNs also fall short.} Another approach to mitigate the impact
of convergence is to leverage SDNs. In existing SDNs, a logically
central controller manages a network of programmable switches. The
controller detects failures, centrally computes forwarding rule
changes, and pushes new rules to switches. However, this approach cannot 
be used to build a fabric which is always available.
First, the controller must
learn about the failure from network switches, which can incur high
latency depending on the placement of the controller in the
network. This can be a factor in Software-defined wide-area
networks (SDWAN)~\cite{b4, swan}. Second, after the controller has
been notified of a failure scenario 
and it has computed global rule changes for multiple traffic classes in
response. Implementing these changes in a running network is challenging. The
controller may have to update the rules of multiple switches using complex
update schedules so that intermediate network states do not lead to
inconsistencies like packet loops and drops~\cite{decentralizedupdate, kinetic,
updatesynthesis}. 
State-of-art SDN update mechanisms can take around \emph{300ms to order of
  minutes} to compute and install the update across a network. Further, He et.
  al~\cite{controlplanelatency} measure the latency for programming rules in
  OpenFlow switches: it can take \emph{10-100}ms to add/modify/delete a single
  rule in the OpenFlow switch tables and such switch rule delays, mainly due to
  inefficient switch control plane software, make consistent updates even
  slower.  Overall, even with SDNs, packets encountering failed links may be
  dropped for extended periods of time until failure notification and consistent
  update installation have completed.


\subsection{Policy Compliance Challenges}

In addition to ensuring availability under failures, we desire
that packets always obey policies: routing around to avoid failure to
reach a destination should not violate policies that pertain either to
communicating with that destination or to network resource management. 

Local failover mechanisms like Fast-Reroute, which are coupled with
distributed routing protocols, cannot guarantee that failover paths
comply with policies.

Even without failures, implementing network-wide policy compliant routing in the network control plane is difficult 
(the underlying problem is NP-Complete in
general~\cite{genesis}.)
In SDN, the centralized controller has to compute new sets of
policy-compliant paths for different flows (identified by packet
headers), which can take \emph{order of seconds to hours}. It has to update
the network in a manner that even the intermediate steps comply with
the policies---this complicates an already arduous process of
generating consistent updates. Determining the appropriate distributed
control plane configurations where high-level policies are met (using
techniques such as~\cite{fibbing,propane,synet,zeppelin}) is even
harder~\cite{zeppelin}. 

\subsection{Our Position}

In sum, policy-compliance while rerouting to ensure always-availability
is difficult to achieve today. An always available fabric in itself is also
hard to achieve.


We argue that the main underlying problem is that a bloated and/or
remote control plane is involved in route computations and/or forwarding
rule installations under failures. We argue for stripping the
control plane down of {\em any} failure reaction duties and
refactoring said duties. In particular we advocate: (1) pushing
recomputation of, and forwarding along, policy-compliant alternate
routes entirely to switch dataplanes, where fastest failure reaction
can happen, and (2) leveraging a logically central policy plane to
allow programming policies and informing switches of what policies to
adhere to for different flows.





\begin{figure}
	\centering
	\includegraphics[width=\columnwidth]{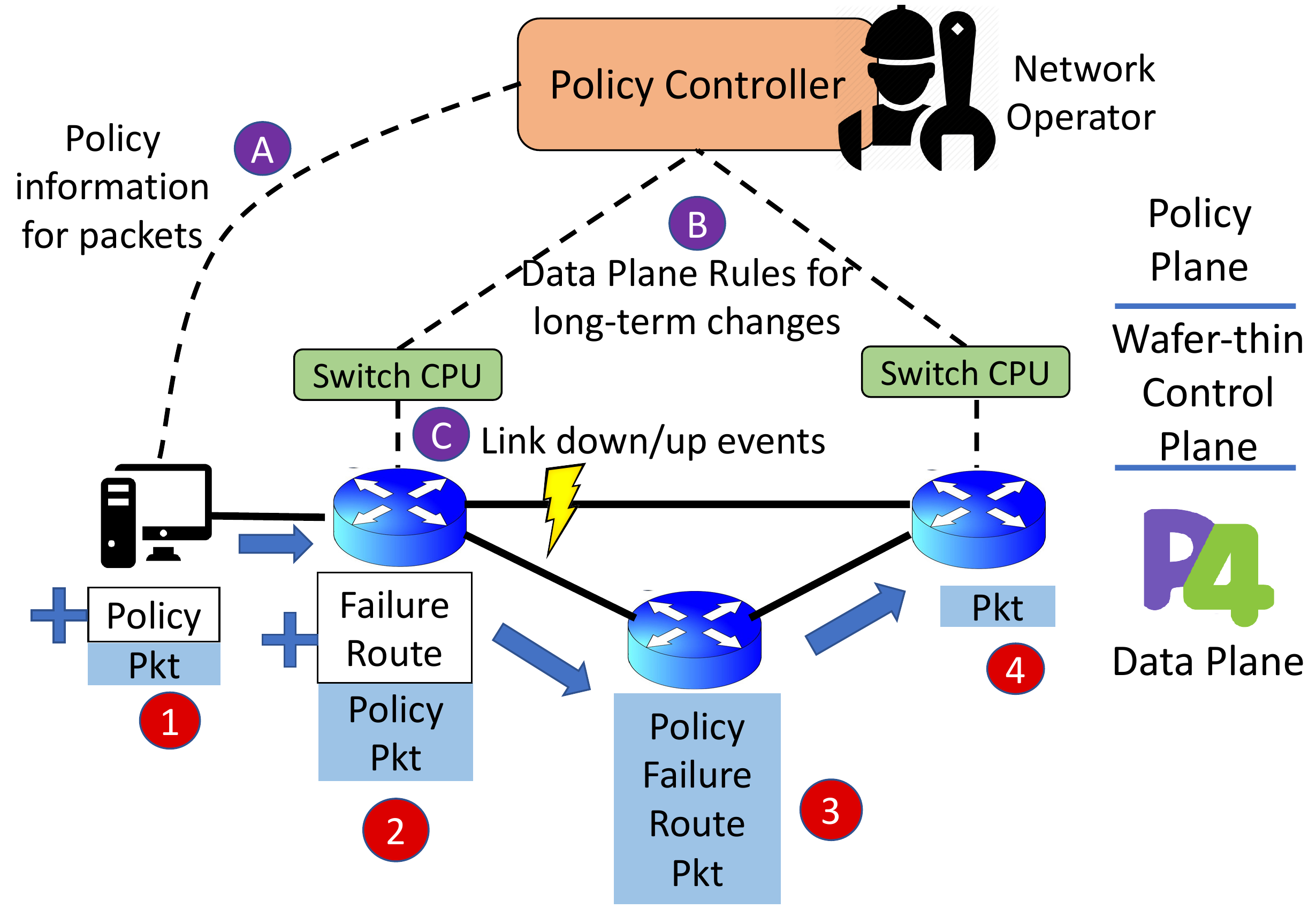}
	\caption{D2R architecture}
	\label{fig:d2rarch}
\end{figure}

\begin{table*}[thb]
\begin{small}
	\begin{center}
		\begin{tabular}{ m{7em} | m{15em} | m{31em} } 
			Policy & API & Description \\
			\hhline{|=|=|=|}
			Middlebox \newline Chaining &
			addMboxChain( flow f, switch[] m1, switch[] m2 ...) 
			& Chain of middlebox arrays where one middlebox is traversed in each array
			Can be coupled with BFS/IDDFS.
			\\ \hline
			Next-hop \newline Preference 
			& addPreference( flow f, switch n1, switch n2)
			& From switch n1, prefer next hop n2 if n1 $\rightarrow$ n2 is active. Can be coupled only 
			with IDDFS.
			\\ \hline
			Weighted Cost \newline Load Balancing 
			& addWeightedLB( flow f, switch n, switch[] next, int[] weights) 
			& At switch $n$, choose next-hop next[1] with probability $\frac{weights[1]}{\sum weights}$ ...
			Can be coupled only with IDDFS. 
		\end{tabular}
	\end{center}
	\vspace*{-3mm}
	\compactcaption{\name Packet Policy Support} \label{tab:policysupport} 
\end{small}
\end{table*}

\section{\name Architecture}\label{sec:architecture}
\name solves the challenges mentioned in \secref{sec:challanges} and
provides always availability and policy-compliance. Key to \name is that
it leverages programmable switch technologies to refactor the
distribution of responsibilities amongst different components of the
network. We illustrate the \name architecture in
\Cref{fig:d2rarch}. In \name, the network is divided into three components:
\begin{compactenumerate}
	\item \textbf{Policy Plane}: The centralized policy controller
          is used by operators to specify the {\em network topology}
          and {\em policy requirements} for different flows. The
          policy plane sends data plane rules to the switch control
          planes, where the rules encode the network's topological  structure
          (\purplec{B}). The policy plane also sends the policy to the
          end-hosts which are then included in the packet headers
          (\purplec{A}).
	\item \textbf{Control Plane}: Our switch control plane is
	{\em wafer-thin}~\cite{waferthincontrolplane} and is mainly responsible for
	programming the data plane with the rules reflecting topology sent by the policy plane 
	(\purplec{B}). The control plane is also
	responsible for monitoring link-up events in the switch (\purplec{C}).
	\item \textbf{Data Plane}: The data plane uses programmable
          ASICs to run {\em graph traversal algorithms},
          atop the network topology encoded in the dataplane rules, to
          compute a policy-compliant route for packets entirely in the
          data plane. Routing does not rely on the control plane or
          policy plane on the critical path (\redc{2}, \redc{3} and
          \redc{4}). For routing under failures, the data planes {\em
            encode link failure information in the packet header},
          which is used for traversal (\redc{2}). The data plane does
          not store global link failure state.
\end{compactenumerate}

We describe the flow of a packet in \Cref{fig:d2rarch}: \redc{1} The packet is
tagged at the end-host with the policy header specified by the policy plane.
\redc{2} When the packet enters the switch, the switch data plane computes a
route through the network taking into account the failed links and the policy, and stores the
route as a source route in the packet. The failure information is also included
in the packet. \redc{3} The switch parses the route in the header and forwards
along the route. \redc{4} The packet is forwarded to the destination. We
outline the roles played by each plane in realizing this forwarding behavior in
the rest of this section. 

\subsection{\name Data Plane}
Modern programmable switching ASICs let developers write complex packet
processing pipelines that can run at very high speeds. 
For instance, the state-of-the-art Barefoot Tofino switch can process packets at an aggregate line rate of
6.5Tbps. Thus, in \name, we move away from the conventional model 
where the
data plane just forwards packets and the control plane runs 
sophisticated routing algorithms. Instead, the data plane runs 
 traversal
algorithms like breadth-first search (BFS) and iterative-deepening depth-first search
(IDDFS) to compute a route from the switch to the destination. Thus, when a packet
arrives at a switch (\redc{2}), the data plane computes a route to the
destination and stores the route in the packet header. Subsequent switches
(\redc{3}, \redc{4}) use the route information in the packet to forward the packet 
to the destination. We describe our P4~\cite{p4}	 implementation of the data plane  in 
\secref{sec:dataplanerouting}. 

Modern programmable ASICs can detect when a connected
link is down and trigger a special packet indicating that the link/port is down
(\purplec{C}). As soon as the failure is detected, the \name data plane stores
this information in a register. When a packet  arrives, the
data plane uses this updated local link-state and computes a route which does
not use the failed link, avoiding any packet drops. 
This approach solves the
problem faced by SDNs, in which failures cause the centralized controller to
react and add new forwarding rules in a consistent manner. Our
approach is more general than static local fast failover mechanisms, as the data
plane can compute a valid next hop dynamically based on the current state of the
links connected to the switch. 

For correct routing in the network, we need to know the state of links
in the entire network. However, we cannot resort to distributed
link-state advertisements because this leads to convergence issues and
packet losses. We eliminate routing convergence periods by using the
\emph{Failure Carrying Packets} (FCP) protocol~\cite{fcp}. In FCP,
each packet carries information about all the link failures it has
encountered in its path (\redc{2}).  The switch data plane uses this
information to find a route that avoids failed links without actually
storing the current global link-failure state. FCP provides guarantees
of connectivity under failures without the need for a distributed
routing protocol. We describe the FCP protocol and implementation in
\secref{sec:fcp}.

\subsection{\name Control Plane}

The control plane in \name is distributed across switches, and it has
a minimal role. Most importantly, the switch control plane plays no part in
the critical path for end-to-end forwarding, and thus, it is not a
bottleneck for always availability and is not responsible for
policy-compliance under failures. The switch control plane simply programs
the data plane with rules provided by the policy plane. These are {\em
  not forwarding rules} and instead they encode the network topology
and any changes that occur to it in the long term, such as planned
maintenance, or links/switches getting (de)commissioned
(\purplec{B}). The policy plane, describe below, tracks these aspects
of the network topology.


Some modern ASICs may not generate a packet for when the link has come back up.
For such scenarios, the switch control plane uses mechanisms like BFD~\cite{bfd}
to monitor the status of links and notify the switch data plane of link up
events (\purplec{C}). 

\subsection{\name Policy Plane}
\name provides support for switch and network-wide policies 
under different failure scenarios. We restrict our
support to \emph{per-packet policies} in the data plane, 
i.e., computing a packet's route is independent from other packet routes. To 
support hyperproperties (a policy constraining 
the routing behavior of two or more flows), 
we would need to store routing state of different flows in the data plane, 
which would consume scarce switch memory resources.

Even for per-packet policies, we need to store the policy information
for different flows. We could store the policy state in the switches,
but if we needed to change the policies, we would need to reprogram
switches, which can lead to down time
(\secref{sec:challanges}). 
Moreover, unlike planned maintenance and link/switch additions/removals that induce slow topology churn, policy
churn is significantly higher and can trigger frequent expensive network
updates. Instead, we develop a policy plane which sends the policy
information to end-hosts (\purplec{A}) which are responsible for
adding the policy in the packet header (\redc{1}). The data plane uses
the policy header to generate policy-compliant paths (\redc{2}).

The policy plane can also request the current state of
network links from switch control planes to generate new policies.
Crucially, for the policies \name supports, policy updates will not
trigger reprogramming of the data plane.  We describe \name's policy
support in \Cref{tab:policysupport} and the data plane implementation
in~\secref{sec:policy}.


\section{Data plane routing} \label{sec:dataplanerouting}
Given the FCP header in the packet, the \name data plane computes an active path
without going to the control plane. In this section, we first present a primer
on programmable switches and P4, the state-of-art language used to program these
switches. We then present two graph traversal algorithms we implement in \name:
breadth-first search (BFS) and iterative-deepening depth-first search (IDDFS).

\subsection{Programmable Switches and P4}
\begin{figure}
	\centering
	\includegraphics[width=0.9\columnwidth]{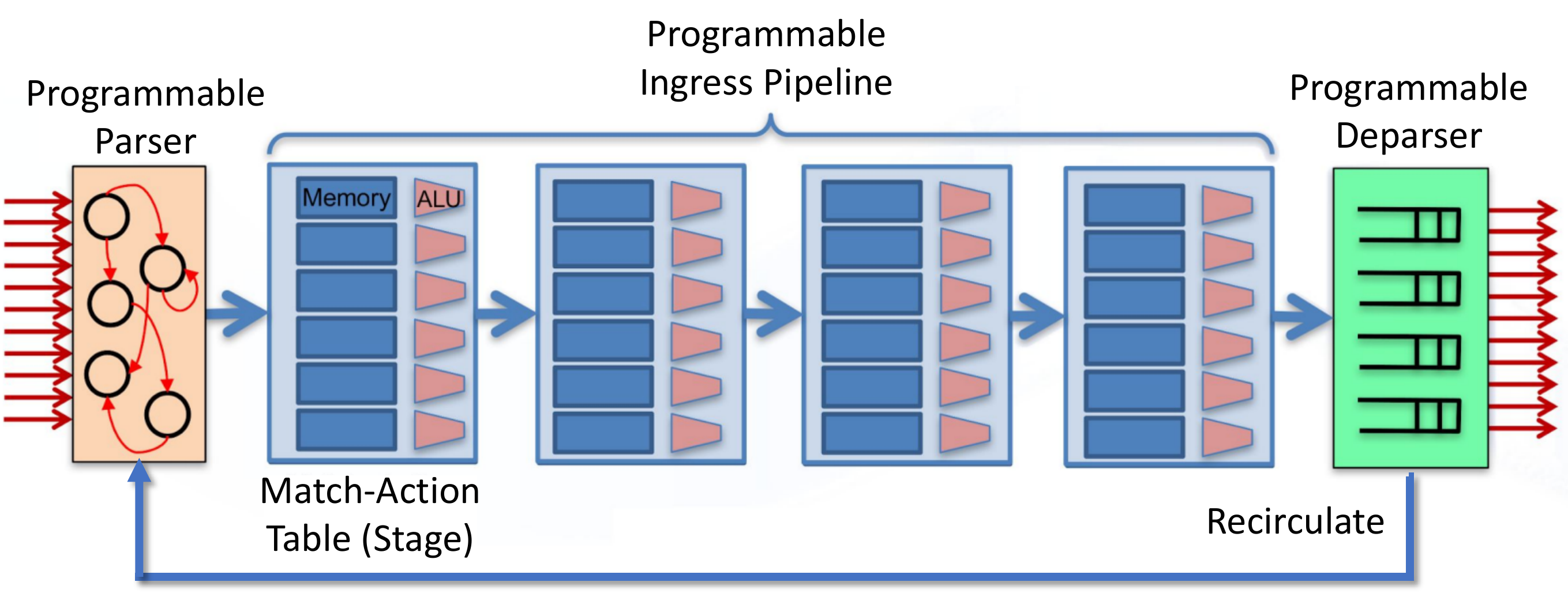}
    \caption{Ingress pipeline in programmable switches}
	\label{fig:ingress}
\end{figure}

Modern programmable switching ASICs~\cite{rmt} contains three main components:
the ingress pipeline, the traffic manager, and the egress pipeline. A switch can have
multiple ingress and egress pipelines serving multiple ingress and egress ports.
Packet processing is performed primarily at the ingress pipelines
(\Cref{fig:ingress}) which comprises of three programmable components: a parser,
a match-action pipeline, and a deparser. To support complex packet processing, each
pipeline has multiple stages which process packets in a sequential fashion. Each
stage contains dedicated resources (e.g., match-action tables and registers)
to process packets at high rates. For instance, the
state-of-the-art Barefoot Tofino switch can process packets at an aggregate line
rate of 6.5Tbps. 

Packet processing can be abstracted as a control flow graph of match-action
tables, where each table matches a set of header fields, and performs 
actions based on the match results. While processing a packet, the stages of the
ASIC share the packet header and metadata fields (can be thought of as global
memory), and stages can pass information in the pipeline by modifying these
headers. The number of stages in programmable switches is limited, and the
packet processing logic may not finish at the pipeline. In such scenarios, the
packet can be \emph{recirculated} back into the ingress pipeline with updated
headers for further processing. Recirculating a packet multiple times
consumes switch bandwidth resources (ports are set up in loopback mode for
recirculations and cannot be used for physical links) and results in increased 
latency. Thus, our data plane algorithms must reduce to a minimum	 the number 
of recirculations required for packet processing.

\begin{figure}
	\centering
	\includegraphics[width=0.5\columnwidth]{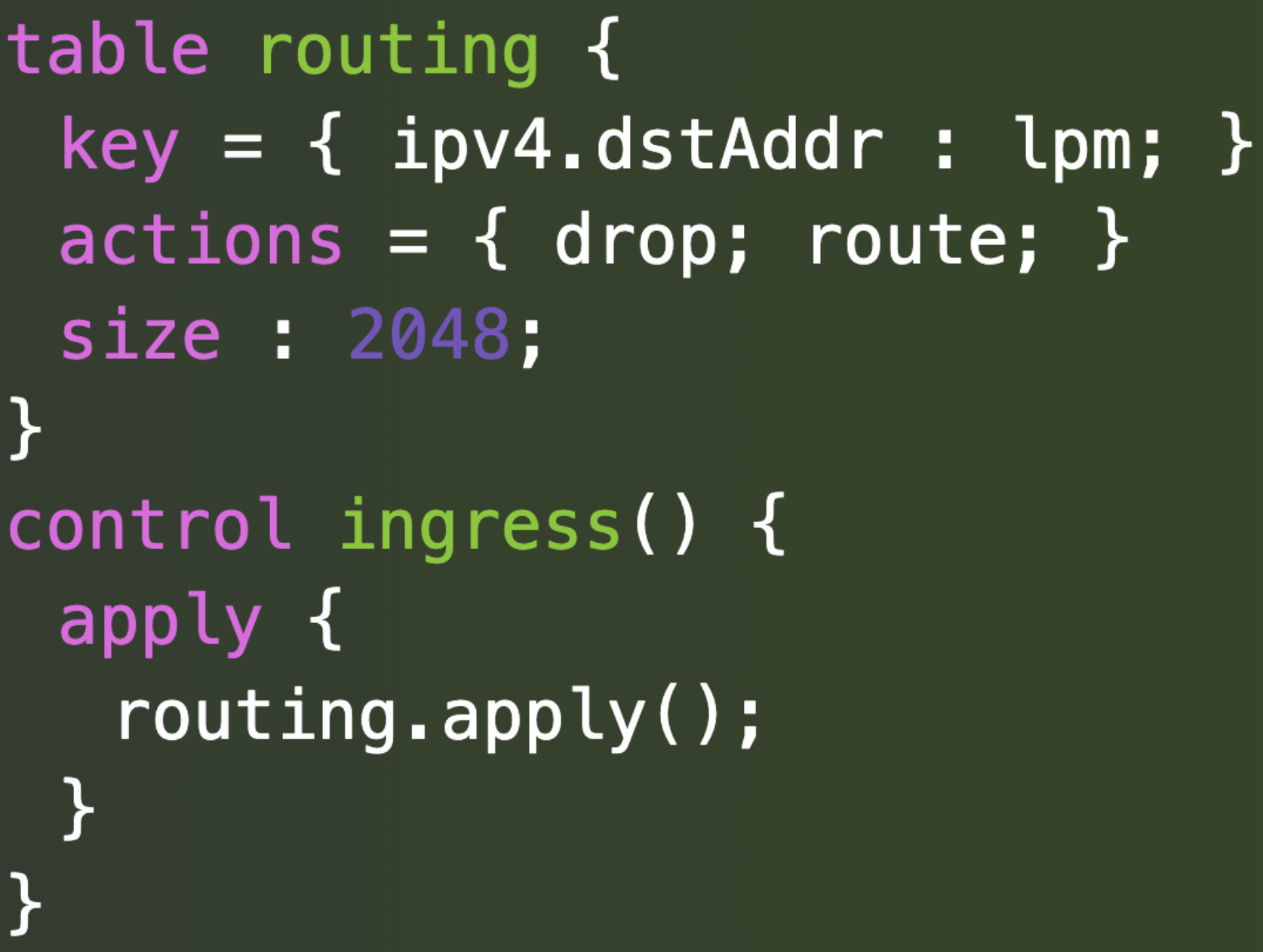}
    \caption{Example P4 program for IPv4 forwarding}
	\label{fig:p4example}
\end{figure}

P4~\cite{p4} is the most widely used domain-specific language to program
these ASICs. \Cref{fig:p4example} illustrates a simple P4 program that defines 
an IPv4 routing table and how the table is invoked in the ingress pipeline. 
While P4 is a programming language, it closely mimics the architecture of programmable 
ASICs---i.e., we cannot express any general algorithm as a P4 program. Thus, we need 
to take into account the P4 semantics for designing our graph traversal algorithms and 
express steps of the routing algorithms as match-action tables.


\subsection{Breadth First Search}
We now present the algorithm and P4 implementation for performing breadth-first
search (BFS) in the network. BFS has the advantage of
finding paths with the least number of hops. Traditionally, BFS explores the switches
of the graph using a first-in-first-out (FIFO) queue. However,
since currently P4 only supports stack data structures, we implement a modified BFS
algorithm in P4 which uses only stacks and preserves the following invariant:
\emph{a switch at a lower depth (number of hops from the source) is explored before any switch at a higher depth}. 
The only 
difference from a queue-based implementation will be 
the relative ordering of explored switches at each depth. We
present our stack-based BFS algorithm in~\Cref{alg:bfs} and in the rest of the section.


\begin{figure}[H]
    \vspace*{-7mm}
	\begin{minipage}{\linewidth}
        \begin{algorithm}[H]
                \footnotesize
				\caption{Stack-based Breadth First Search}
				\label{alg:bfs}
				\begin{algorithmic}[1]
                    \Procedure{BFS}{src, dst}
                    \State{Initialize array of Stacks[MaxDepth]}
                    \State{depth = 0}
                    \State{curr = src}
                    \While {curr != dst and Stack array is not empty}
                        \For{next in Neighbors(curr) \label{alg:neighbor}} 
                            \If{(curr, next) is not visited or failed}
                            \State{// Add valid neighbor to stack of depth + 1}
                            \State{Stack[depth + 1].push(next) \label{alg:nextdepth}}
                            \State{Mark all incoming edges to next as visited \label{alg:neighboractionend}}
                            \State{Parent[next] = curr}
                            \State{\textbf{goto While}}
                            \vspace*{1.5mm}
                            \EndIf
                        \EndFor
                        \If{Stack[depth] is not empty \label{alg:stack}}
                            \State{// Explore next switch at current depth}
                            \State{curr = Stack[depth].pop();\label{alg:stackend}}
                        \Else
                            \State{// Explored all switches at current level. Move to depth + 1}
                            \State{depth++ \label{alg:switch}}
                            \State{curr = Stack[depth].pop();}
                        \EndIf
                    \EndWhile
                    \vspace*{1.5mm}
                    \State{Traverse Parent map from dst$\rightarrow$src to compute path}
					\EndProcedure
				\end{algorithmic}
		\end{algorithm}
    \end{minipage}
    \vspace*{-7mm}
\end{figure}

\paragraph{P4 implementation.}
In programmable switches, the amount of memory to store packet headers and
metadata is limited. Since the BFS stacks need to be processed by every stage,
we need to store it as a header field\footnote{We will not emit these stacks in
the deparser as they are not required for correct forwarding in the network.},
thus, we must limit the number of used stacks. 
Our BFS algorithm
uses two stacks for odd (Stack[1]) and even depth (Stack[0]) switches,
respectively--- when we are exploring switches of odd depth $d$, we  push the
neighbors at depth $d+1$ in Stack[1] and vice-versa, eliminating the need of more
than two stacks. 

\begin{figure}
	\centering
	\includegraphics[width=\columnwidth]{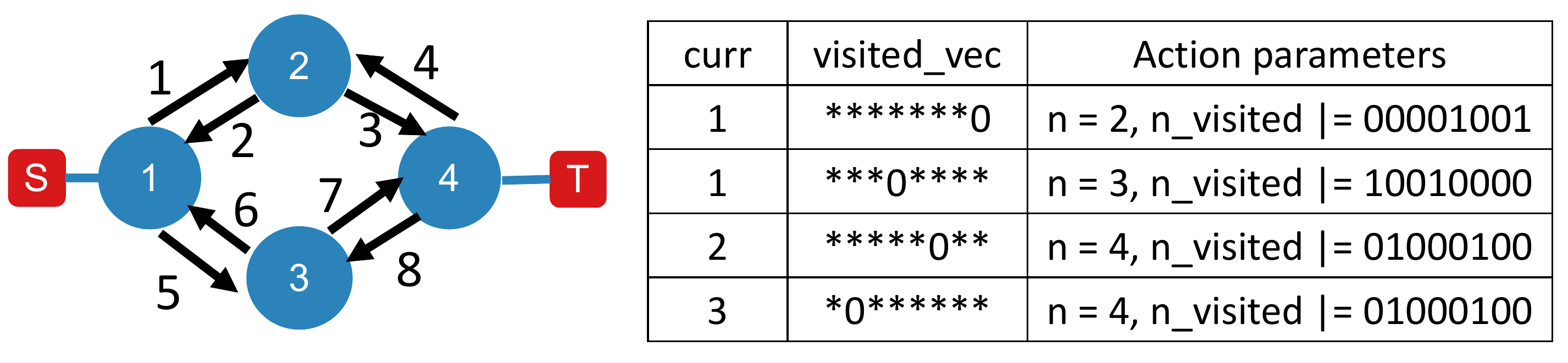}
    \caption{Example topology of 4 switches and the subset of
    BFS table rules.}
	\label{fig:bfsexample}
\end{figure}

We now break down how we translate \Cref{alg:bfs} 
to P4. The building block of our BFS algorithm is the following P4 match-action table:
\begin{Verbatim}[frame=single]
table bfs { key={
    hdr.curr : exact;
    hdr.visited_vec : ternary;
    hdr.stack: exact;
  } actions =
  {push_neighbor; pop_stack; change_stack;}
\end{Verbatim}

\paragraph{Initialization.} 
We initialize \texttt{curr} to the switch that is computing a path to the
destination. \texttt{visited_vec} is a bitvector whose size is equal to the
number of bidirectional links. For each $link_i$,
\texttt{visited_vec[i]}\footnote{The indices start from 1 from the rightmost bit
of the vector.} is set to 1 if $link_i$ has been visited or has failed, and to 0
otherwise. We set all failed links obtained from the FCP header to $1$. Consider
the example in \Cref{fig:bfsexample}. If $1 \rightarrow 2$ has failed, then we
set the $1^{st}$ and $2^{nd}$ bits of \texttt{visited_vec}---0000 0011. We also
set all incoming links to \texttt{curr} to 1, so that BFS does not visit
\texttt{curr} later in the algorithm. For our 2-stack implementation, we denote
switches at odd depth with \texttt{stack} = 1, and switches at even depth with
\texttt{stack} = 0.

Let $m$ be a switch at odd depth. BFS explores a neighbor $n$ which is
unvisited and connected (line~\ref{alg:failcheck}) by an active link and puts in
Stack[1] (line~\ref{alg:nextdepth}). To map our algorithm into the  
P4 programming model, we translate the if condition to a table match rule, and the 
code executed based on the if condition as one of the table actions. 

If the link ID of $m \rightarrow n$ is $id$, then we will only explore $n$ from
$m$ if \texttt{visited_vec[id]} = 0. Consider the example in
\Cref{fig:bfsexample}. If $m=1$, we will only explore $1 \rightarrow 2$ if
\texttt{visited_vec[1] = 0}; P4 supports ternary match kind for bitvectors where
we can specify exact values (0/1) or wildcard for each bit; we use the ternary
match to check the $id^{th}$ bit in \texttt{visited_vec}. Thus, the match fields
for exploring the edge $m \rightarrow n$ would be as follows (depending on if
$m$ is at odd or even distance from source):
\begin{Verbatim}[frame=single]
  curr: m=1, visited_vec: *******0, stack:0
  curr: m=1, visited_vec: *******0, stack:1
\end{Verbatim}
If the above match succeeds, we need to push $n = 2$ onto Stack[1]. We also set
all the bits corresponding to incoming edges to $n$ as 1 in
\texttt{visited_vec}; thus, BFS will not explore $n$ again. For this, we define
the following action which implements
lines~\ref{alg:nextdepth}-\ref{alg:neighboractionend}
\footnote{P4 targets may not support specifying header fields as indices--- we define 
two action push_neighbor_0 and push_neighbor_1 to push onto Stack[0] and Stack[1] respectively.
We elide these details for simplicity.}.
\begin{Verbatim}[frame=single]
action push_neighbor(n, n_visited) {
  Stack[~hdr.stack].push(n);
  hdr.visited_vec = hdr.visited_vec|n_visited;}
\end{Verbatim}
\Cref{fig:bfsexample} shows the action parameters when we explore 
the edge $1 \rightarrow 2$.
Once, all neighbors of $m$ are explored, the BFS algorithm will pop the next
element from the stack of the current depth (odd or even) and repeat the process
of exploring the neighbors (lines~\ref{alg:stack}-\ref{alg:stackend}). To check
if all neighbors of $m$ have been explored, we again use the ternary match to
check if all bits corresponding to outgoing links of $m$ are 1. If
so, we update \texttt{curr} to the top switch of the stack. For example, if
$m=1$, the links with ID 1 and 5 must be explored:
\begin{Verbatim}[frame=single]
match{
curr: m=1, visited_vec:***1***1, stack:0
curr: m=1, visited_vec:***1***1, stack:1
}
action pop_stack() {
    hdr.curr = Stack[hdr.stack].pop();}
\end{Verbatim}

Finally, once we have explored all switches in the stack, we need to proceed to the
switches at the next level. To match for this condition, we place a special switch "0" at
the bottom of stack and swap stacks when once \texttt{curr} = 0. After
switching stacks, we pop the top element of the new stack to start exploring its
neighbors.
\begin{Verbatim}[frame=single]
match{
curr:0, visited_vec:********, stack:0
curr:0, visited_vec:********, stack:1
}
action change_stack() {
    hdr.stack = ~ hdr.stack; 
    hdr.curr = Stack[hdr.stack].pop();}
\end{Verbatim}

\paragraph{Ingress implementation.} 
According to the P4 semantics, only one match-action rule will be triggered per
table application. The match condition depends on the current packet headers,
priorities and ordering of rules in the table. In the ingress pipeline, when a
table is applied, the switch will execute the action code corresponding to the
matched rule. Thus, a single \texttt{bfs} table application cannot perform the
entire traversal. We apply multiple \texttt{bfs} tables to
perform BFS from the source till \texttt{curr} is the destination switch. 

\begin{Verbatim}[frame=single]
control ingress {
    bfs_1.apply();
    if (hdr.curr != hdr.dst) {bfs_2.apply();}
    else {forwarding.apply();}
    if (hdr.curr != hdr.dst) {bfs_3.apply();}
    else {forwarding.apply();}
    ...
    // End of pipeline
    if (hdr.curr == hdr.dst) {
        forwarding.apply();
    } else {
        recirculate();}}
\end{Verbatim}

Each \texttt{bfs} table reads and writes the packet headers (\texttt{curr,
visited_vec, stack}) which are passed down in the pipeline to the next \texttt{bfs}
table. Therefore, there is a Read-After-Write (RAW) dependency between the bfs
tables. Thus, they cannot be placed in the same stage~\cite{compiling-dependencies}. 
Switches
only have a bounded number of stages ($\sim$10) in the ingress pipeline. Therefore, our
BFS algorithm may not reach the destination in those stages. To overcome the
limitation of bounded number of stages, we can repeatedly \emph{recirculate} 
the packet back into the ingress pipeline
with the headers at the end of the pipeline. This effectively resumes the BFS
algorithm, and we keep applying the \texttt{bfs} table till we find the
destination in the algorithm. To avoid recirculations, we implement a source 
routing flavor of BFS--- the route is stored in the packet headers and 
downstream switches can use the source route and avoid route recomputations (thus, 
recirculations). We also propose a hierarchical routing scheme in ~\secref{sec:hierarchy} 
to further reduce recirculations by splitting route computations across switches.

\begin{figure}
	\begin{minipage}{\linewidth}
		\begin{algorithm}[H]
                \caption{Iterative Deepening Depth First Search}
                \footnotesize
				\label{alg:dfs}
				\begin{algorithmic}[1]
                    \Procedure{DFS}{src, dst}
                    \State{Initialize empty Stack}
                    \State{curr = src}
                    \State{len = 0}
                    \State{max_len = 4}
                    \While {curr != dst}
                        \If{len < max_len}
                            \For{next in Neighbors(curr)}
                                \If{(curr, next) is not failed or visited \label{alg:failcheck}}
                                \State{// Go to valid neighbor \label{alg:nexthop}}
                                \State{Mark all incoming edges to next as visited}
                                \State{Stack.push(curr)}
                                \State{Parent[next] = curr}
                                \State{curr = next}
                                \State{len = len - 1 \label{alg:nexthopend}}
                                \State{\textbf{goto While}}
                                \EndIf
                            \EndFor
                            \vspace*{1.5mm}
                            \State{// No valid neighbor. Backtrack}
                            \State{curr = Stack.pop() \label{alg:btvisited}}
                            \State{len = len - 1}
                        \Else
                            \State{// current length exceeds max length. Backtrack}
                            \State{curr = Stack.pop() \label{alg:btdepth}}
                            \State{len = len - 1}
                        \EndIf
                        \vspace*{0.3mm}
                        \If{curr == NULL and Stack is empty \label{alg:resetstart}}
                            \State{// Explored all switches within max_len distance} 
                            \State{// Increase max_len exponentially}
                            \State{max_len = max_len $\times$ 2}
                            \State{curr = src}
                            \State{Reset visited state \label{alg:resetend}}
                        \EndIf
                    \EndWhile
                    \vspace*{1.5mm}
                    \State{Traverse Parent map from dst$\rightarrow$src to compute path}
					\EndProcedure
				\end{algorithmic}
		\end{algorithm}
	\end{minipage}
\end{figure}

\subsection{Iterative Deepening Depth First Search}
Another form of graph traversal that can yield paths while exploring fewer
number of switches (thus, fewer recirculations) is Depth-first Search (DFS).
However, without bounds on the path length, DFS can produce very long paths 
compared to
BFS. This is not ideal, especially in wide-area settings. 
We implement a variant of DFS called Iterative Deepening DFS (IDDFS), which explores switches in a
manner similar to DFS while imposing bounds on the length of the discovered paths, and iteratively
increases the bound when needed.
We present our IDDFS algorithm in~\Cref{alg:dfs}.

IDDFS works similarly to DFS with one major modification: we keep track of the length
of the current path from src (\texttt{len}) and will not explore neighbors if
the length of the path exceeds the max length path. 
Thus, IDDFS provides bounds on the path length and will eventually find a path
if one exists within the bound. 
If a path within the bound does not exist, we perform a new DFS  with an increased bound. 
IDDFS is linear in complexity. In the worst case, it explores $2N$ switches.

\paragraph{P4 Implementation.} Similar to BFS, we create a P4 table 
which acts as the building block of our IDDFS algorithm.
\begin{Verbatim}[frame=single]
table iddfs {
  key = {
    hdr.curr : exact;
    hdr.visited_vec : ternary;
    hdr.len: exact;
    hdr.max_len: exact;
  } actions =
  {goto_neighbor; backtrack; increase_length;}
  default action = backtrack();
\end{Verbatim}
Similar to BFS, we add table rules 
to check if certain edges are 
visited/failed (using ternary match) and explore neighbors. Backtracking occurs
when we have no neighbor to visit from a switch. Finally, we increase the 
maximum path length when the stack is empty---i.e., when we have explored all 
switches at the specified maximum length but did not reach the destination. 
The P4 Implementation details are in\secref{sec:dfsp4}.

As with BFS, each invocation of the 
\texttt{iddfs} table can lead to one action execution. 
Thus, we add $n$ tables staged one after the other (due 
to the RAW dependency). At the last stage, if we have not 
found the destination, we recirculate the packet again. 
Similar to BFS, we implement source routing for IDDFS. 
IDDFS \emph{requires} source routing for \emph{correctness} purposes 
as it does not compute the shortest path to the destination. 
Consider the topology in \Cref{fig:bfsexample}. Switch 1 
uses IDDFS and computes the route $1\rightarrow 2 \rightarrow 4$ 
(but does not store it in the packet) and sends to switch 2. 
Switch 2 now performs IDDFS to compute route $2 \rightarrow 1 \rightarrow 3 \rightarrow 4$,
and sends it back to 1, and thus, packet will keep oscillating. 
Oscillation is circumvented by source routes: switch 2 will simply 
use the source route to send to 4.



\section{Hierarchical Routing} ~\label{sec:hierarchy}

In the BFS and IDDFS algorithms we presented in \secref{sec:dataplanerouting},
the computation and memory requirements on each switch increase as the 
network size increases. 
First, increased memory requirements lead to complex resource
fitting problems on the switch. 
Second, and most important, an increased number of traversal computations 
leads to \textit{more recirculations} that consume precious switch capacity.
In \secref{sec:zooeval},
we measure the
recirculation incurred by IDDFS for a network with 126 links and
observe that IDDFS routing can incur up to 10 recirculations to compute routes. 
This limitation of BFS and IDDFS routing begs the question: 
\vspace*{-1mm}
\begin{center}
\textit{Can we avoid computing paths
for the entire network and decrease the number of recirculations?}
\end{center}
Our approach is inspired by OSPF's idea of dividing
the network into areas to avoid large link-state databases on routers.
In the rest of the section, we present \textit{\name
hierarchical routing}, a routing mechanism that reduces recirculation overhead. 

\begin{figure}
	\centering
	\includegraphics[width=0.55\columnwidth]{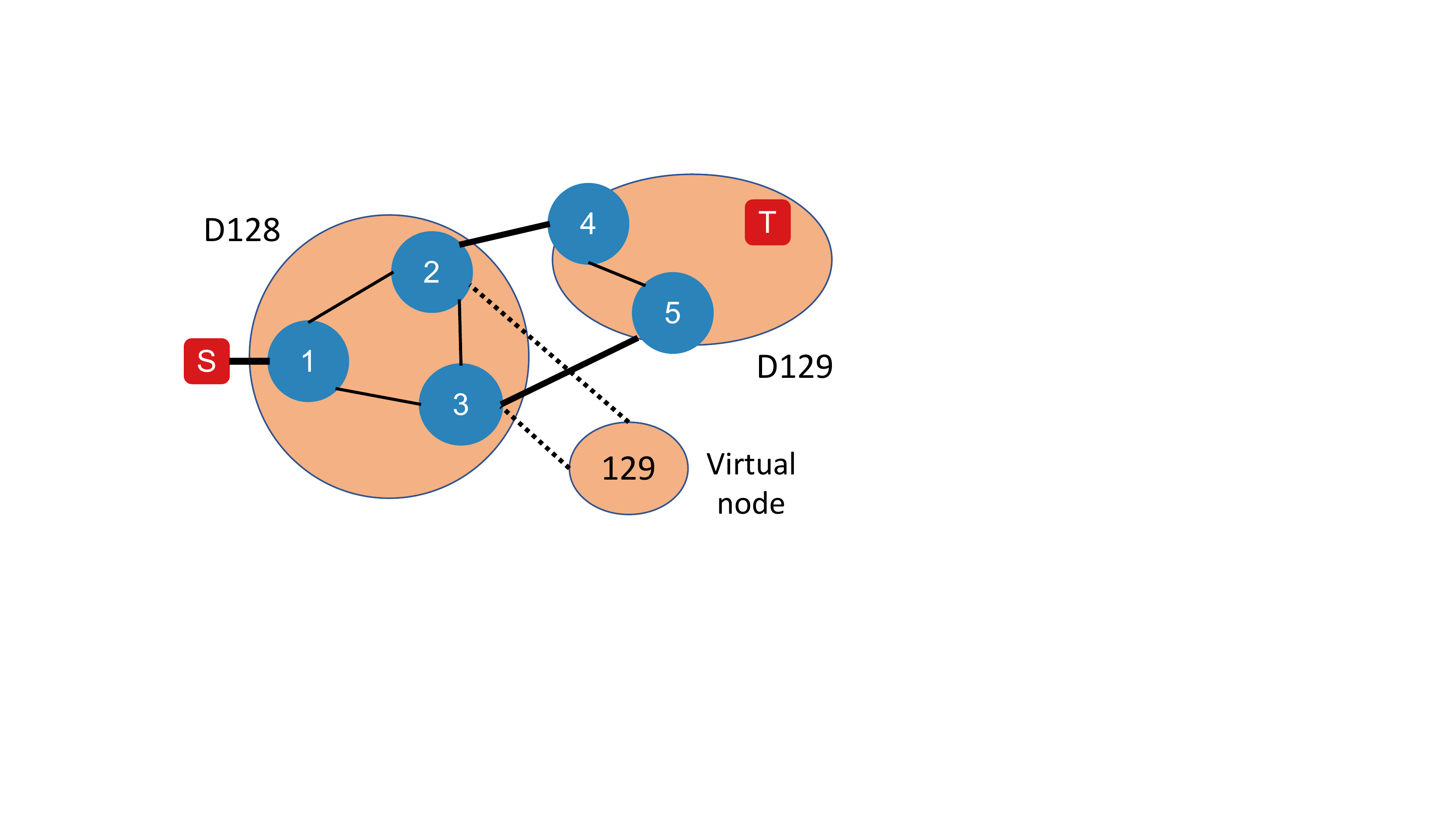}
    \caption{Example of hierarchical routing.  For domain 128, we add
    virtual switch 129 and add links corresponding to $2-4$ and $3-5$.}
	\label{fig:hierarchyex}
\end{figure}

\paragraph{Routing Across Domains.} 
We divide the network into $n$ domains and construct a domain graph based on the
domain adjacencies---i.e., if there is a switch $n_1$ of domain $d_1$ connected
to switch $n_2$ of domain $d_2$, we add an edge between $d_1$ and $d_2$ in the
domain graph. 
Hierarchical routing works as follows: 
(1) The source switch computes the domain path to the destination domain and stores
the path in the packet header.
(2) The source switch then computes the intra-domain network
path to a switch that belongs to the next domain in the domain path, and sends
the packet to that domain, and so on till we reach the destination domain.
(3) The switch in the destination domain finds a path to the destination.
In summary, instead of finding the complete network path in a single switch, we split
the computation across multiple domains, and the first switch is also responsible
for finding a path in the domain graph. 
We implement the switching logic in our
ingress table actions (details omitted for brevity).

We extend our graph algorithms to perform a traversal over the domain graph and
store the domain path in the header, which is then used by the switches to find
a path through each of the domains. We use the BFS/IDDFS tables defined in
\secref{sec:dataplanerouting} for finding  both the domain path and the network
path, and differentiate between the two modes using a header field in the table
match conditions: \texttt{hdr.hierarchy} = 1 means we are finding a domain path,
and \texttt{hdr.hierarchy} = 0 means we are finding a path inside the domain. 

\paragraph{Routing Inside Domains.} 
A switch has to find a route to one of the switches in the next domain. We
modify the topology of each domain to add a special switch for each neighboring
domain. We take all inter-domain links and connect them to the special domain
switch. We illustrate this augmentation in \Cref{fig:hierarchyex}. Thus, to find a
path to the next domain $d$, we set \texttt{hdr.destination = d} and perform
BFS/IDDFS on the modified topology---thus, finding a valid path to the next
domain. Consider a packet from S to T in \Cref{fig:hierarchyex}. Switch 1 will
first compute the domain path to T which is $128 \rightarrow 129$. Then, it will
perform intra-domain BFS/IDDFS to 129 in the augmented intra-domain graph and
will reach either switch 4 or 5 (based on if route is computed through 2 or 3,
respectively). Switch 2 and 3 will have forwarding rules to send the packet to 4
and 5, respectively. Once the packet has reached a switch in domain 129, the
switch can perform intra-domain routing to reach the destination. 

\paragraph{Hierarchical Routing under failures.}
We modify the FCP failure vectors to account for inter-domain link 
failures. Consider the example in \Cref{fig:hierarchyex}: the domain link 
128 - 129 can be marked as failed only if both $2-4$ and $3-5$ links have 
failed. Thus, we can create a mapping of the network failvector to domain 
failure vector (implemented using a match-action table)---the domain failure 
vector can be then used to perform traversal on the domain graph. However, 
unlike normal FCP routing,  hierarchical routing does not provide strict 
guarantees of reachability: if a domain becomes internally disconnected, we 
may not find a route to the destination even if one exists. To provide strict 
routing guarantees, we fall back to single domain routing (the whole 
network is a single domain) whenever a switch is unable to find a route using 
the hierarchical routing rules.  

\begin{figure}
	\centering
	\includegraphics[width=0.75\columnwidth]{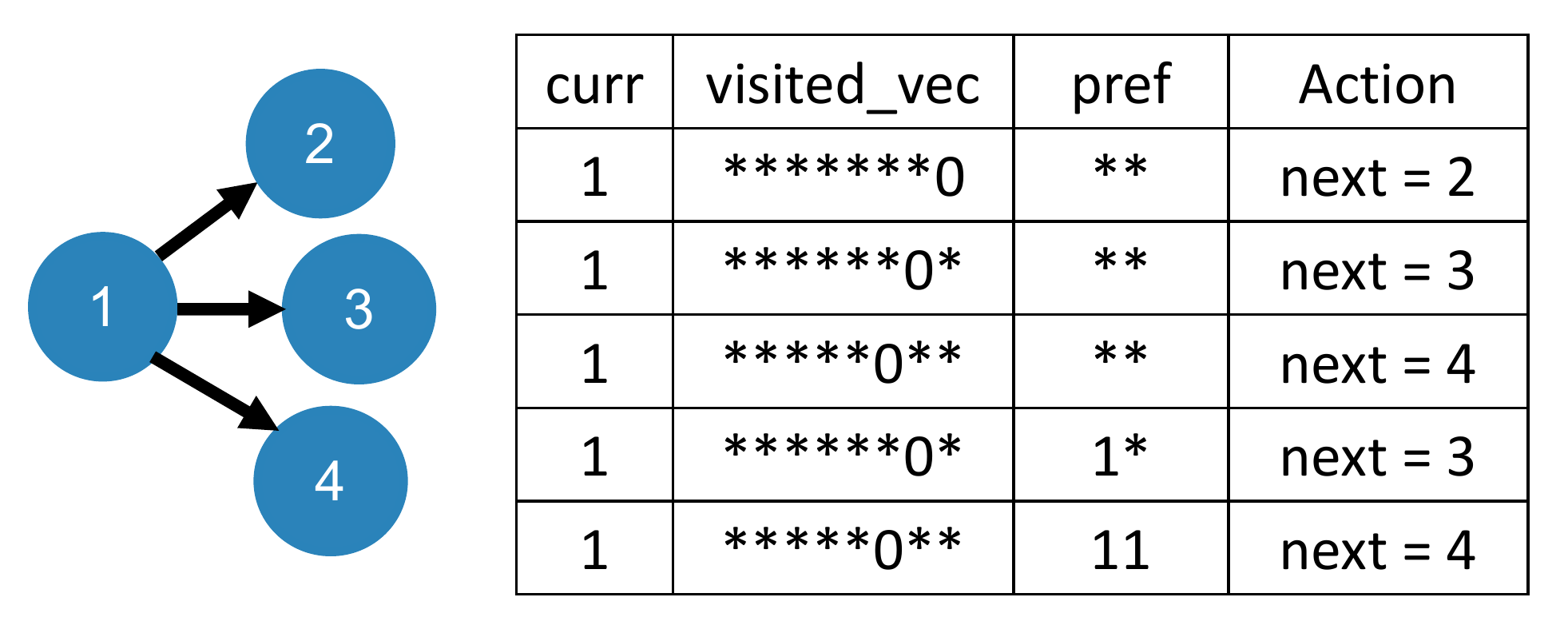}
	\caption{Example preference values for a switch with 3 next hops}
	\label{fig:prefex}
\end{figure}

\section{Policy Implementation} \label{sec:policy}
The \name data plane can find compliant routes for the packet policies listed in
\Cref{tab:policysupport}, even under failures. The operator specifies the
policies to the policy plane using the API, and the policy plane specifies the
policy information which must be sent on the packet, which is used to guide 
the traversal. In this section, we
present the modifications to our IDDFS implementation to support policies. \name 
has the following failure semantics: if there exists a policy-compliant route 
in the network, the data plane algorithm will find it. 

\subsection{Middlebox Chaining}
With the emergence of NFV~\cite{opennf, e2}, operators can place 
middleboxes at different locations in the network to perform
different network functions---e.g., firewall, intrusion detection, 
traffic optimizers etc.,
With the middlebox chaining policy, operators can specify a chain of
middleboxes m1 $\rightarrow$ m2 $\ldots$ and the data plane must compute a
path from src to m1, then to m2 $\ldots$ and then to the destination. The 
middleboxes and destination are encoded in the packet header, and the 
data plane sets \texttt{hdr.dst} = m1, so then IDDFS will find a 
route to m1. Once, the path is found to m1, we set
\texttt{hdr.destination} = m2 and restart traversal from m1, and so on until
the switch computes a route to the destination. The other switches can use the 
computed route for forwarding. Under failures, a switch will be able to 
compute a new route through the middleboxes. 

We also have support for specifying middlebox replicas. With support for
disjunctions in P4 conditional statements, we can end IDDFS when
\texttt{hdr.curr} is equal to any of the middlebox instances. So, we add multiple fields
in the header to store the replicas and modify our ingress pipeline as
follows:
\begin{Verbatim}[frame=single]
control ingress {
  ...
  if (hdr.curr != hdr.dst[1] 
      or hdr.curr != hdr.dst[2]...)
    // apply iddfs
  else 
    // switch to next middleboxes/destination
    // or forward to next-hop        
\end{Verbatim}
Note that enforcing the middlebox policy will not incur any additional 
rules or per-flow state; the policy in the packet header will specify the 
middlebox chain and replicas, which will be read by the data plane.

\subsection{Next-hop Preferences} \label{sec:preferences}
Operators may impose the most preferred path among multiple paths available to a
destination, so that the fabric prefers or avoids using certain paths for cost
or performance reasons. Preferences can be used 
by the operator to send a particular class of traffic through a geographical
domain which has higher bandwidth or is less prone to malicious entities.
\name supports next-hop preferences (akin to BGP
local preferences), which can be used to specify at switch $n$ the best next-hop
$b$ for the packet. To enforce this policy in the data plane, we need to ensure
that when our traversal reaches $n$, it must choose $n \rightarrow b$ if the
link is active and routes to the destination. For next-hop preference, we use
the IDDFS traversal to find a route. In IDDFS, the hop
which is explored first is the most preferred hop (as IDDFS will move to $b$ and
so on till it finds the route to destination), thus, we need to enforce that the
rule $n \rightarrow b$ is matched first in IDDFS. We cannot use rule priorities
as they will require control plane intervention for different policies. 

We add a new longest prefix match (lpm) field to the \texttt{iddfs} table: \texttt{hdr.pref}. 
For each switch and next-hop, the policy plane decides the \texttt{pref}
value to guide IDDFS towards the most preferred hop. 
We illustrate the preferences using an example in \Cref{fig:prefex}. Suppose the
policy specifies that 4 is the most preferred hop from 1, for which the
\texttt{pref} value is set to 11. By virtue of the lpm match, the $5^{th}$ rule
will be the most preferred rule and IDDFS will explore 4. Similarly, if we set
\texttt{pref} = 10, the switch will match to the $4^{th}$ rule and switch 3 will
be the most preferred route. Finally, if we set \texttt{pref} = 00, all $1^{st}-3^{rd}$ 
rules are valid matches with equal length prefixes (**). According to the P4
switch semantics, the first rule will be matched, and IDDFS will explore switch
2. The policy plane is responsible for specifying the right preference value in
the packet depending on the policy, and the data plane will explore the
appropriate hop if it is active. We do not support backup preferences in the
data plane (prefer b1, then b2 etc.). However, if the preferred link is down, we ensure 
we pick an active route (to ensure high availability). 

\subsection{Flexible Weighted Load-Balancing}
One of the key responsibilities of network routing is load-balancing---sending
different flows on different paths to manage network capacity.
\name supports flexible WCMP~\cite{wcmp} 
in the data plane---i.e., the packet will carry the WCMP weights for a switch, and
the switch's data plane will find a route by picking a next-hop with probability
calculated by the weights specified in the packet. The data plane logic does not 
depend on any particular set of weights. Thus, we can simply change weights in the 
packet and the data plane would perform load-balancing according to the new weights.
In current networks,
the control plane needs to add a set of rules based on fixed WCMP weights---if one 
needs to change weights, the control plane needs to modify the data plane, and if 
one of the next-hop links is failed, the switch would drop packets. 

\begin{figure*}
    \centering
	\subfloat[k=0]{\includegraphics[width=0.5\columnwidth]{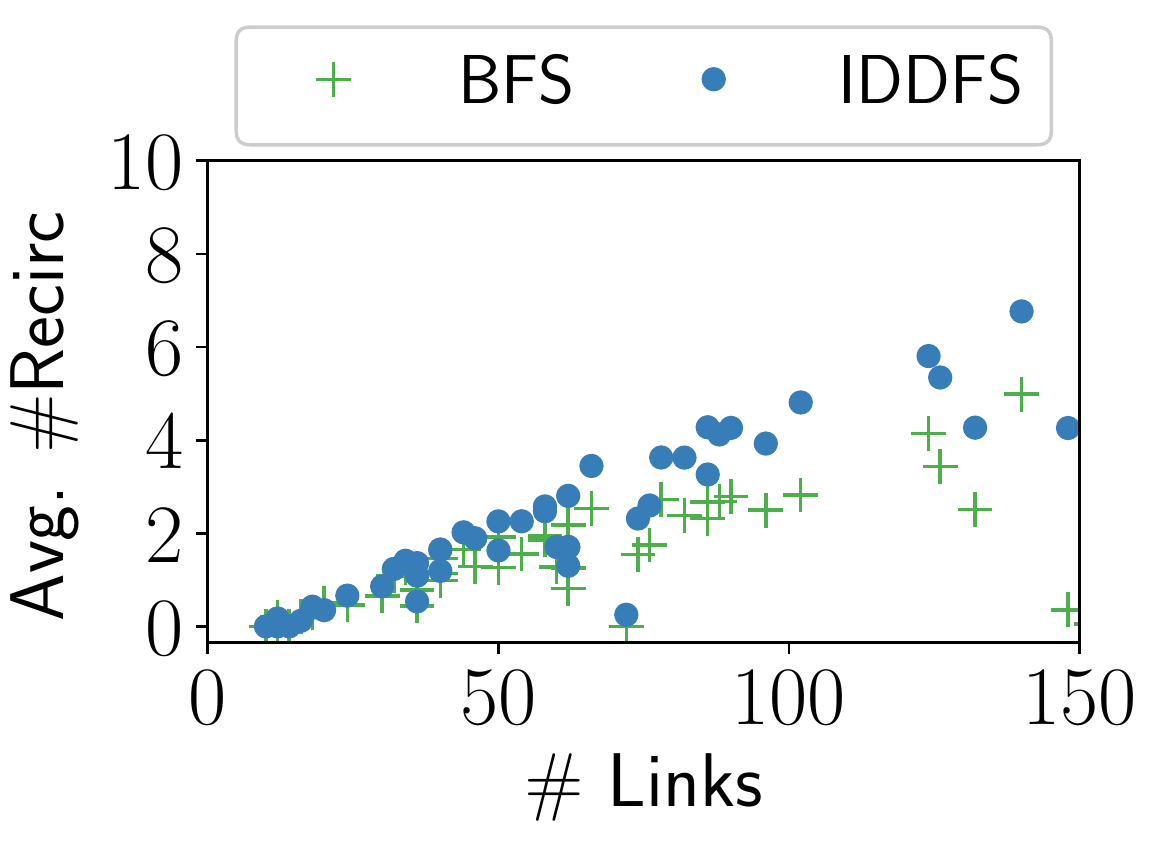}}
    \subfloat[k=1]{\includegraphics[width=0.5\columnwidth]{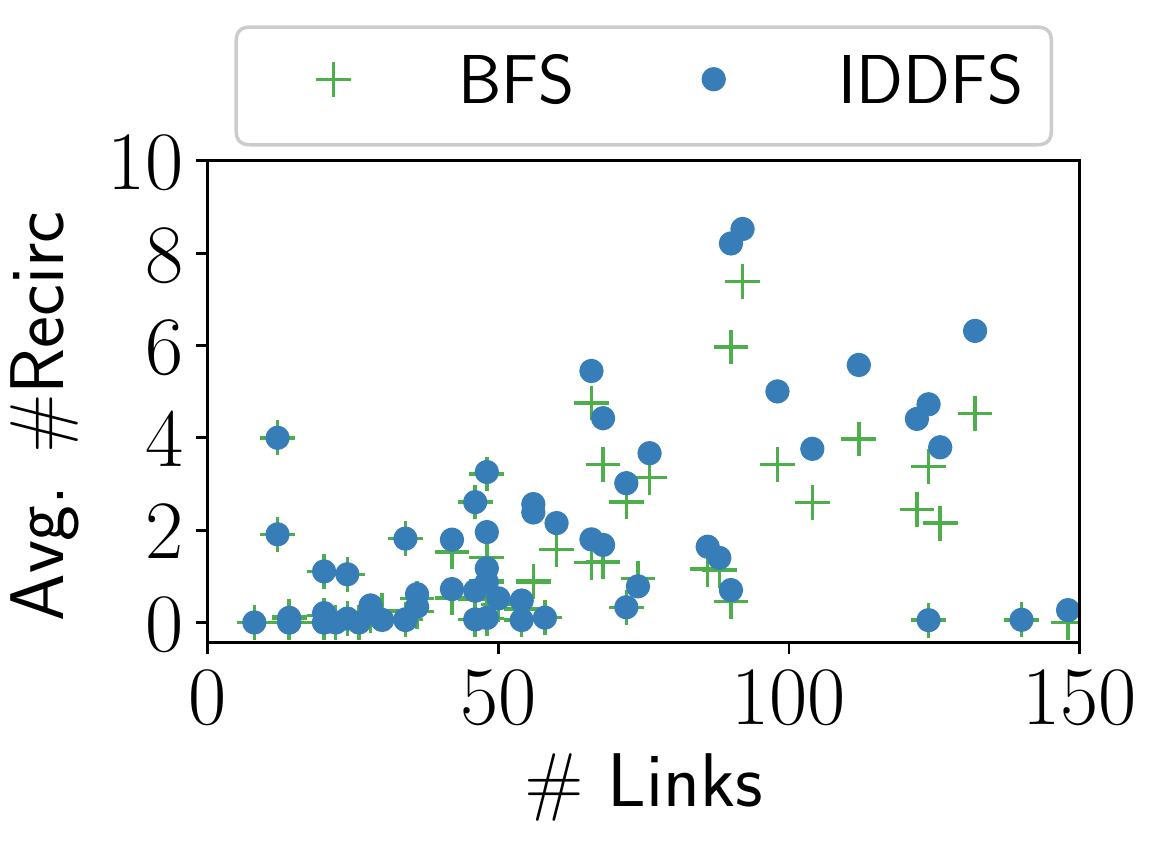}}
    \subfloat[k=2]{\includegraphics[width=0.5\columnwidth]{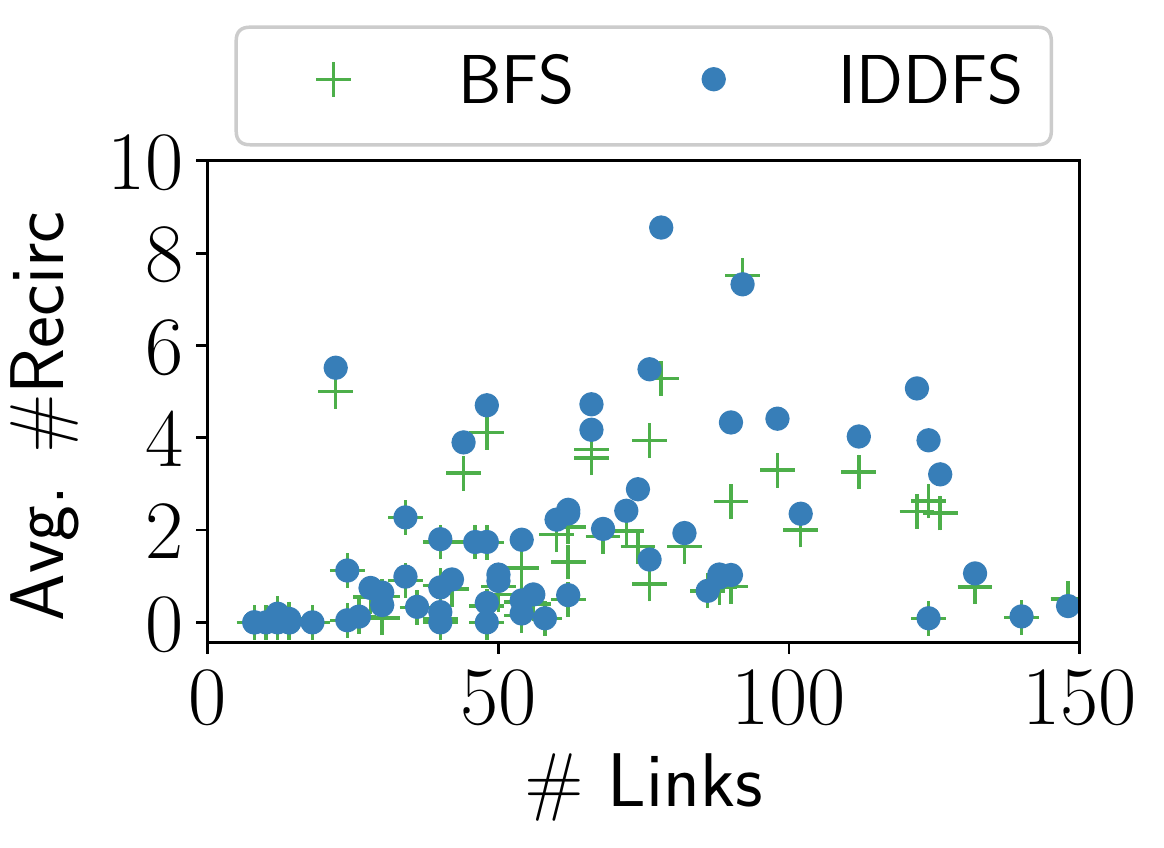}}
    \subfloat[k=3]{\includegraphics[width=0.5\columnwidth]{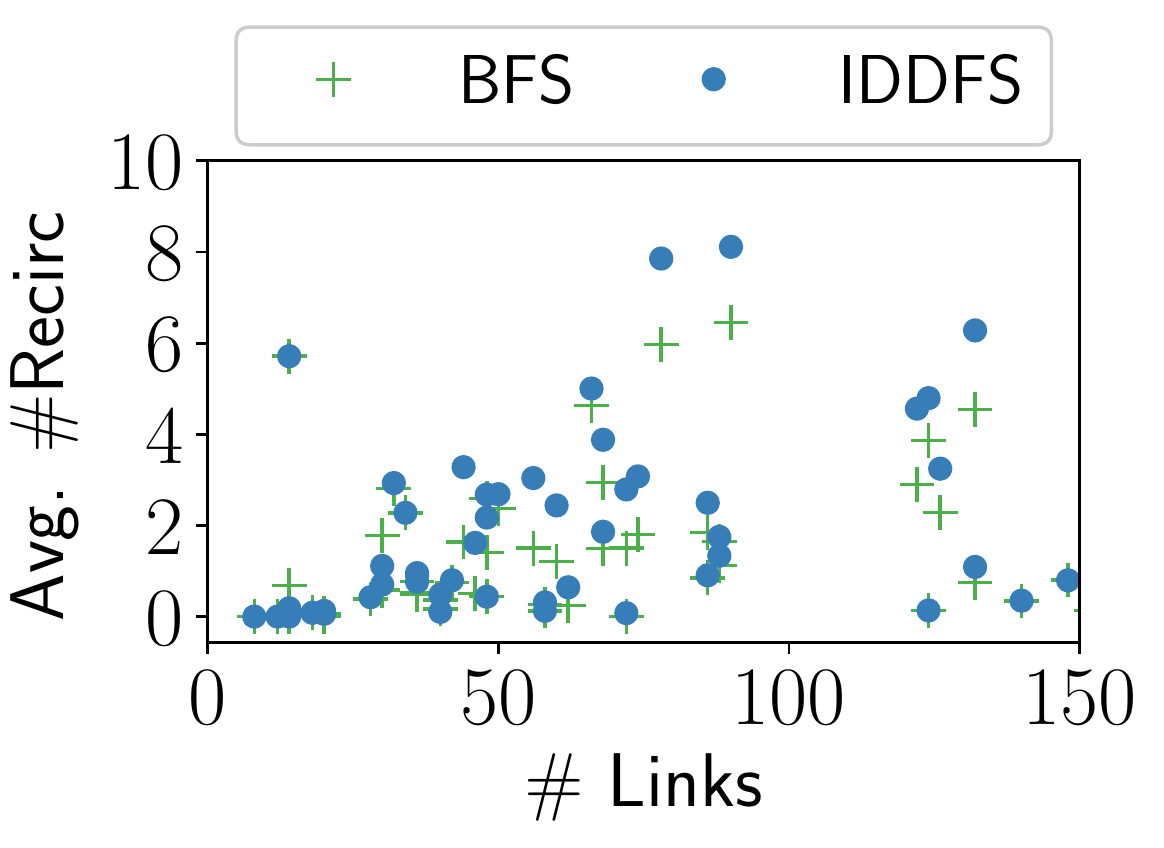}}
    \vspace*{-3mm}
    \caption{\label{fig:recirczoo} \footnotesize
        Average network \# recirculations for varying networks 
        under different $k$-link failure scenarios.}
    \vspace*{-3mm}
\end{figure*}

\begin{figure*}
    \vspace*{-3mm}
    \centering
	\subfloat[k=0]{\includegraphics[width=0.5\columnwidth]{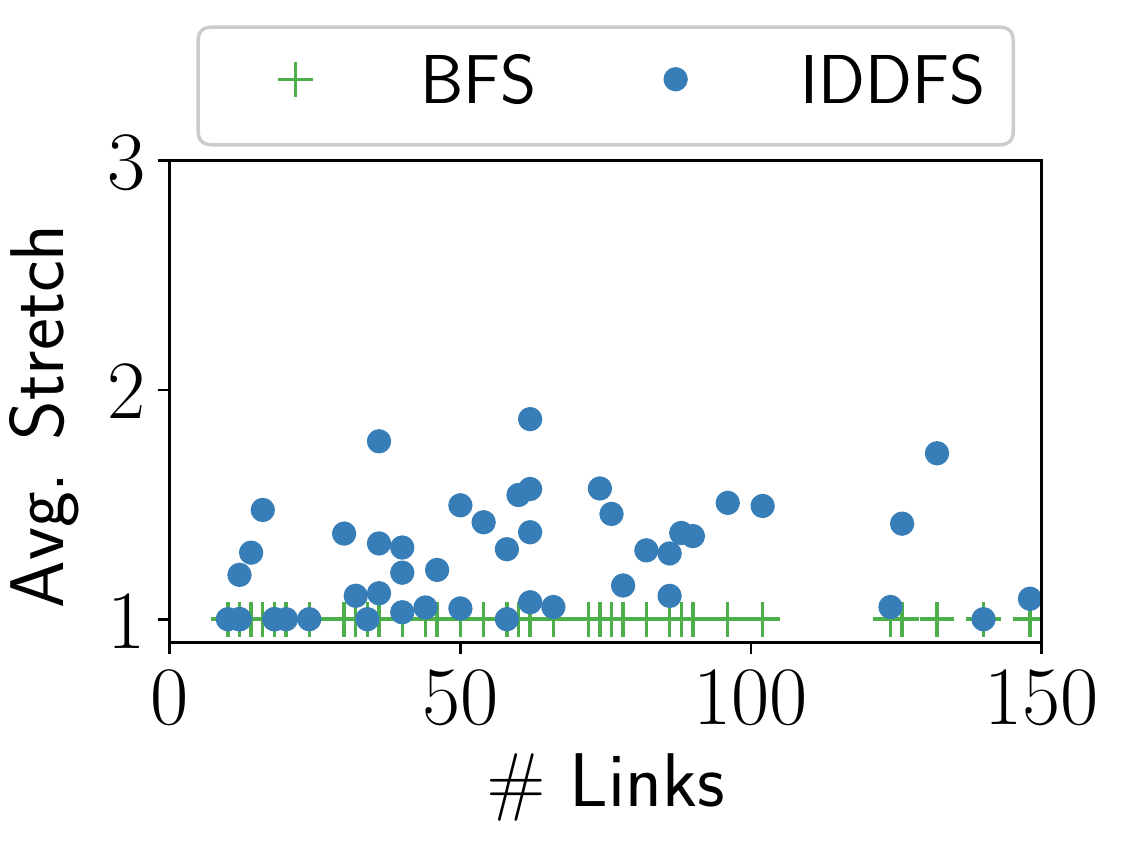}}
    \subfloat[k=1]{\includegraphics[width=0.5\columnwidth]{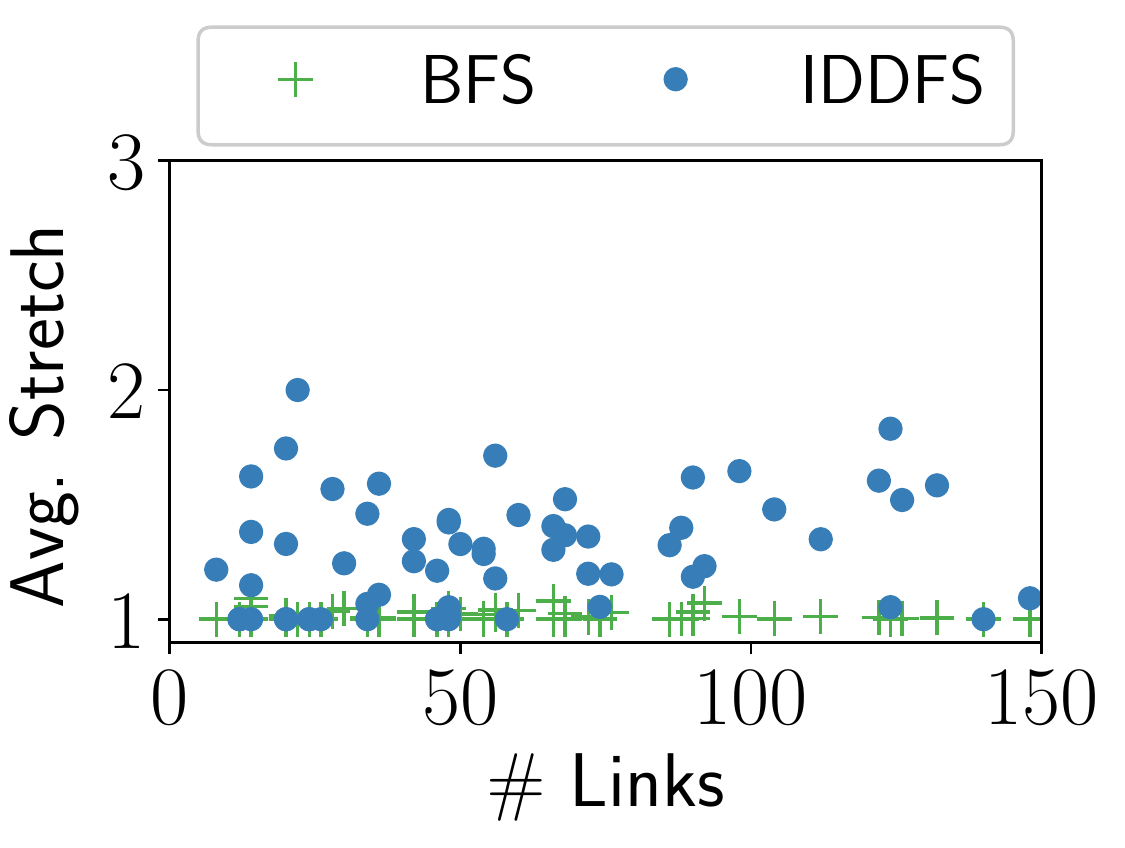}}
    \subfloat[k=2]{\includegraphics[width=0.5\columnwidth]{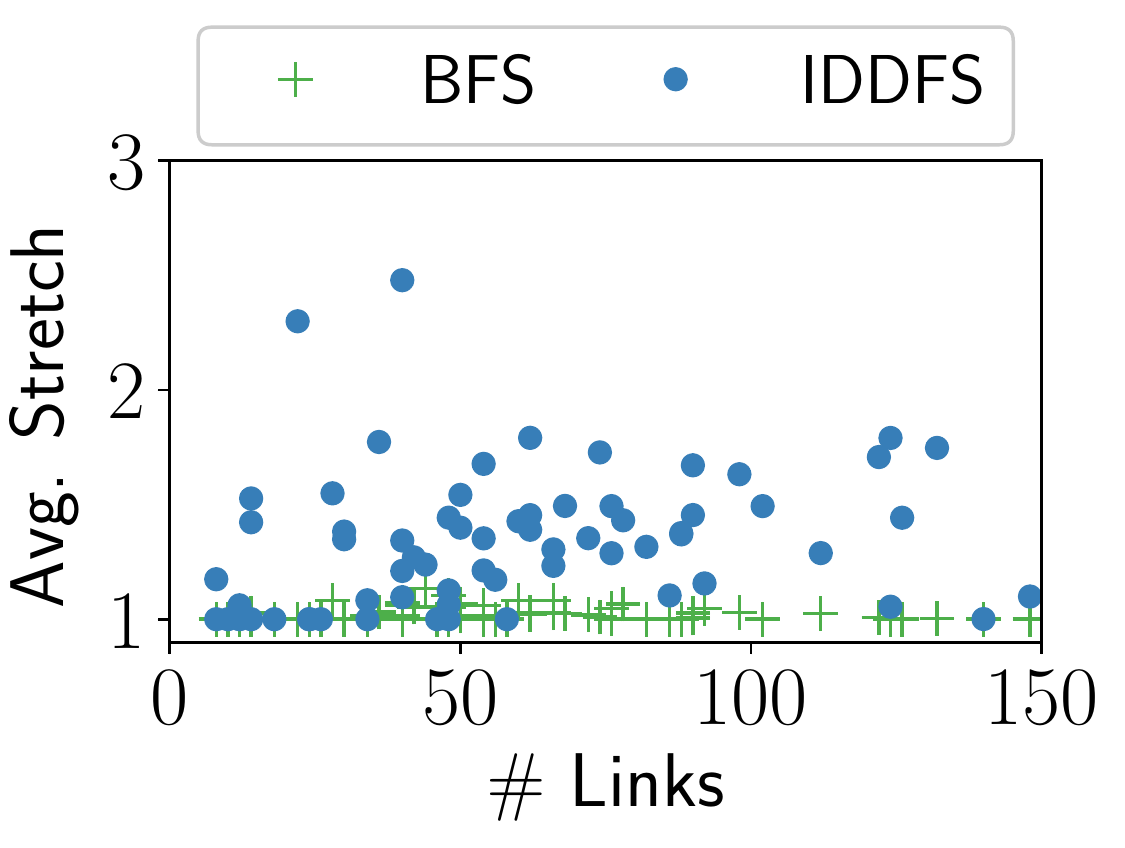}}
    \subfloat[k=3]{\includegraphics[width=0.5\columnwidth]{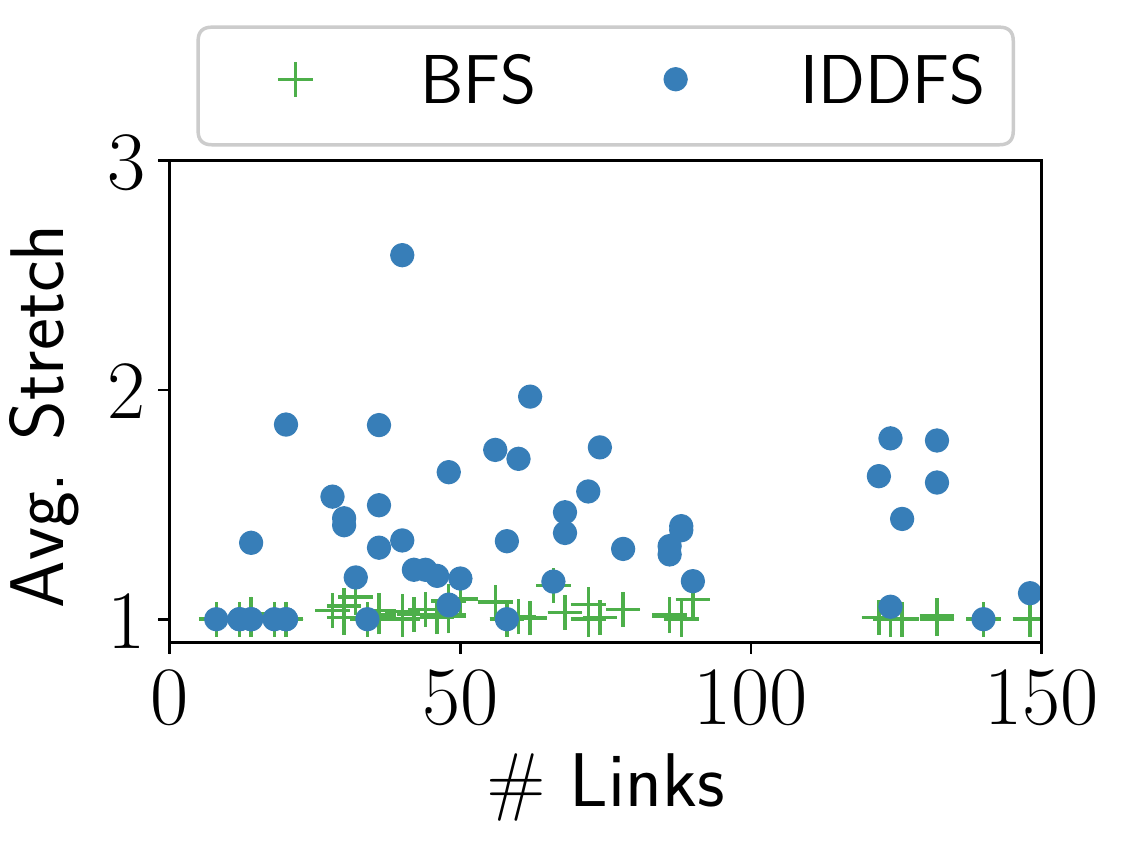}}
    \vspace*{-3mm}
    \caption{\label{fig:stretchzoo} \footnotesize
        Average stretch (ratio of length of path taken vs. shortest path) for varying networks 
        under different $k$-link failure scenarios.}
    \vspace*{-3mm}
\end{figure*}

We illustrate how \name avoids this problem. Consider the switch in
\Cref{fig:prefex}. Assume the policy in the packet specifies load-balancing weights
as 1:2:1. We use preferences presented in \secref{sec:preferences} to load
balance flows according to the weights in the packet. The data plane should
set \texttt{hdr.pref = 00} with probability 1/4 for switch 2, \texttt{10} with
probability 2/4 for switch 3, and finally, \texttt{11} with probability 1/2 for
switch 4. Thus, flows will be load-balanced at switch 1 with weights 1:2:1. P4
switches have support for generating hashes from the packet header fields, which
\name uses to decide the next-hop preference in a probabilistic manner. To
support flexible WCMP, we use Boolean operations in a preprocessing table to map
the random hash to a preference value based on the input weights. In the face of
failures, we prefer a next-hop from the active next-hops with the same relative
weights. For example, if the policy for switch 1 in \Cref{fig:prefex} is 1:2:1
and link $1\rightarrow 3$ is down, the links 
$1\rightarrow 2$ and $1\rightarrow 4$ will be preferred in a 1:1 ratio.
For brevity, we elide the P4 implementation details.

\paragraph{Policy Support Limitations.}
We currently do not support next-hop preferences and weighted load balancing
policies with BFS traversal. BFS explores multiple routes simultaneously, so
choosing one of the BFS routes which comply with the policy requires more
complicated processing in the tables and increased header state, thus, inflating
the number of stages required to find the path (thus, more recirculations). BFS
works in conjunction with middlebox policies.

We can only support limited policies with hierarchical routing. For e.g.,
middlebox traversals cannot be completely enforced as a switch only computes a 
path within a domain---we can specify intra-domain middleboxes. Path preferences 
and load balancing can be supported with hierarchical routing as they only guide 
traversal to pick a particular next-hop. These limitations 
are subject to future work.

\begin{figure}
    \vspace*{-3mm}
    \centering
	\subfloat[\# Recirculations]{\includegraphics[width=0.5\columnwidth]{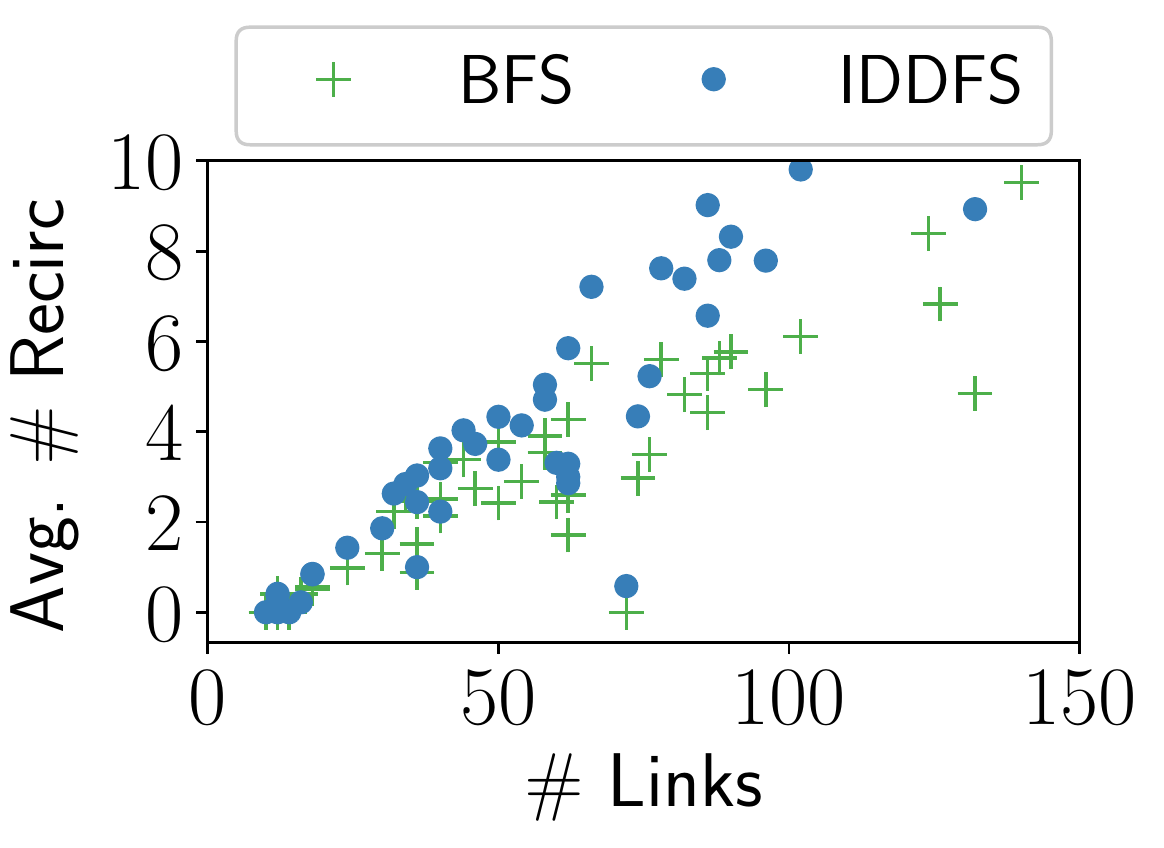}}
    \subfloat[Stretch]{\includegraphics[width=0.5\columnwidth]{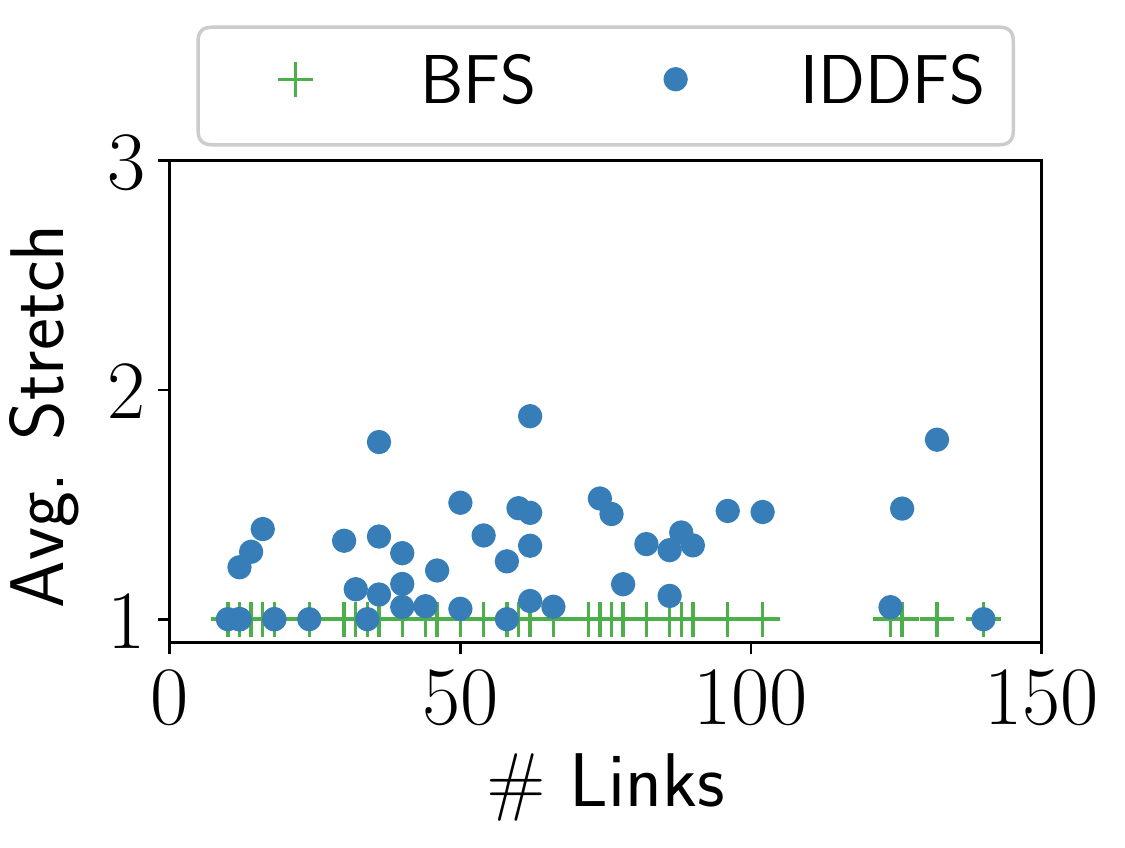}}
    \vspace*{-3mm}
    \caption{\label{fig:middlebox} \footnotesize
        Average \# recirculations and stretch for 100 middlebox policies}
\end{figure}

\begin{figure*}
    \centering
	\subfloat[USA-Recirc]{\includegraphics[width=0.5\columnwidth]{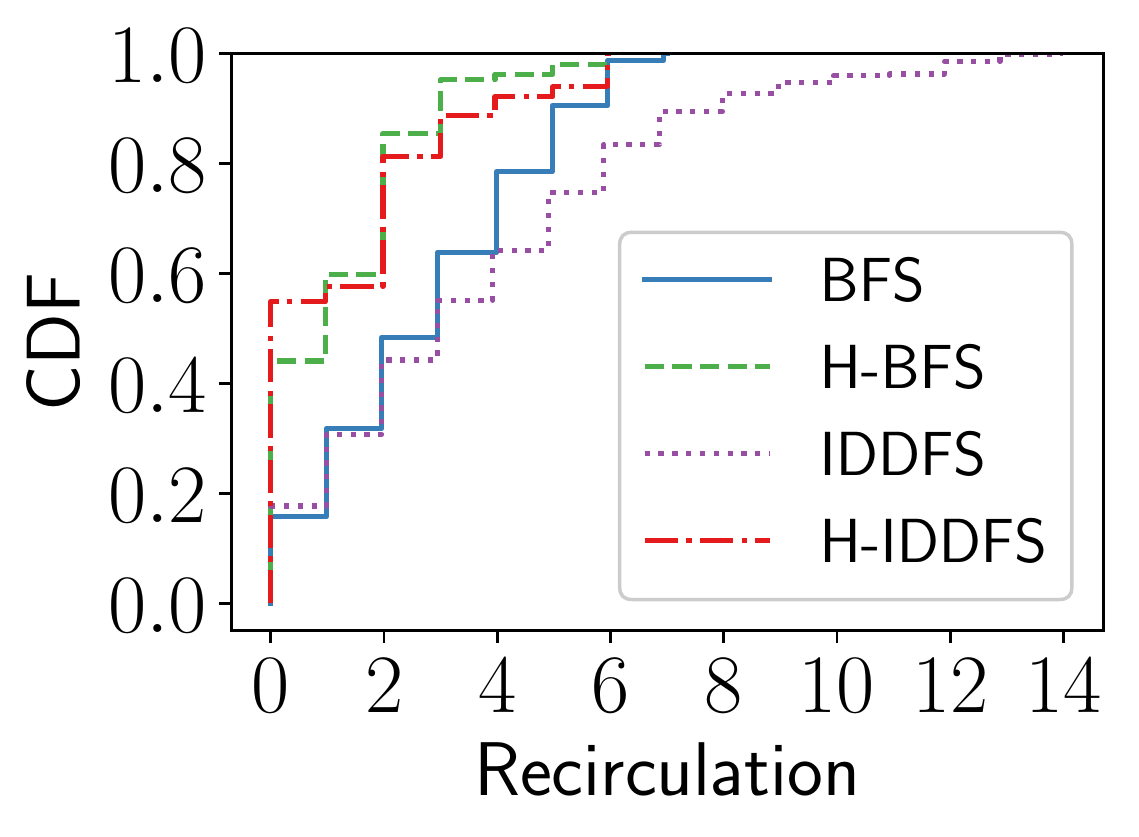}}
    \subfloat[Cesnet-Recirc]{\includegraphics[width=0.5\columnwidth]{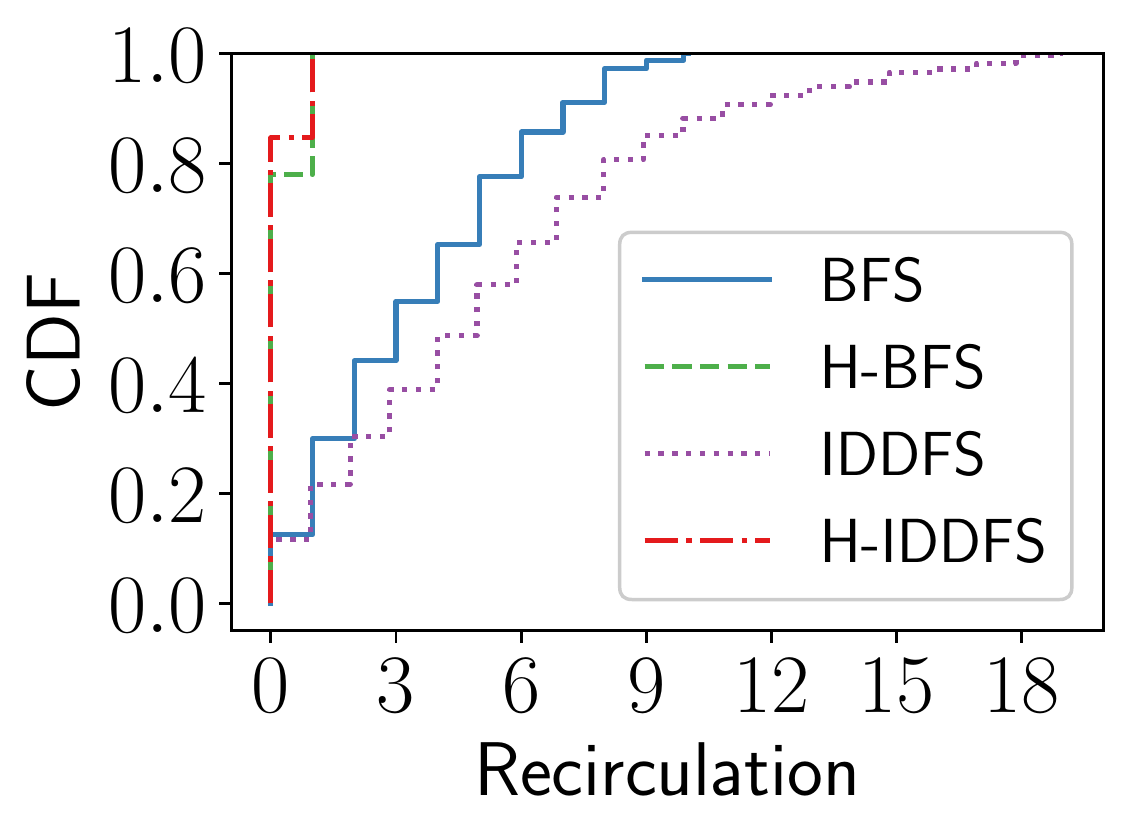}}
	\subfloat[USA-Stretch]{\includegraphics[width=0.5\columnwidth]{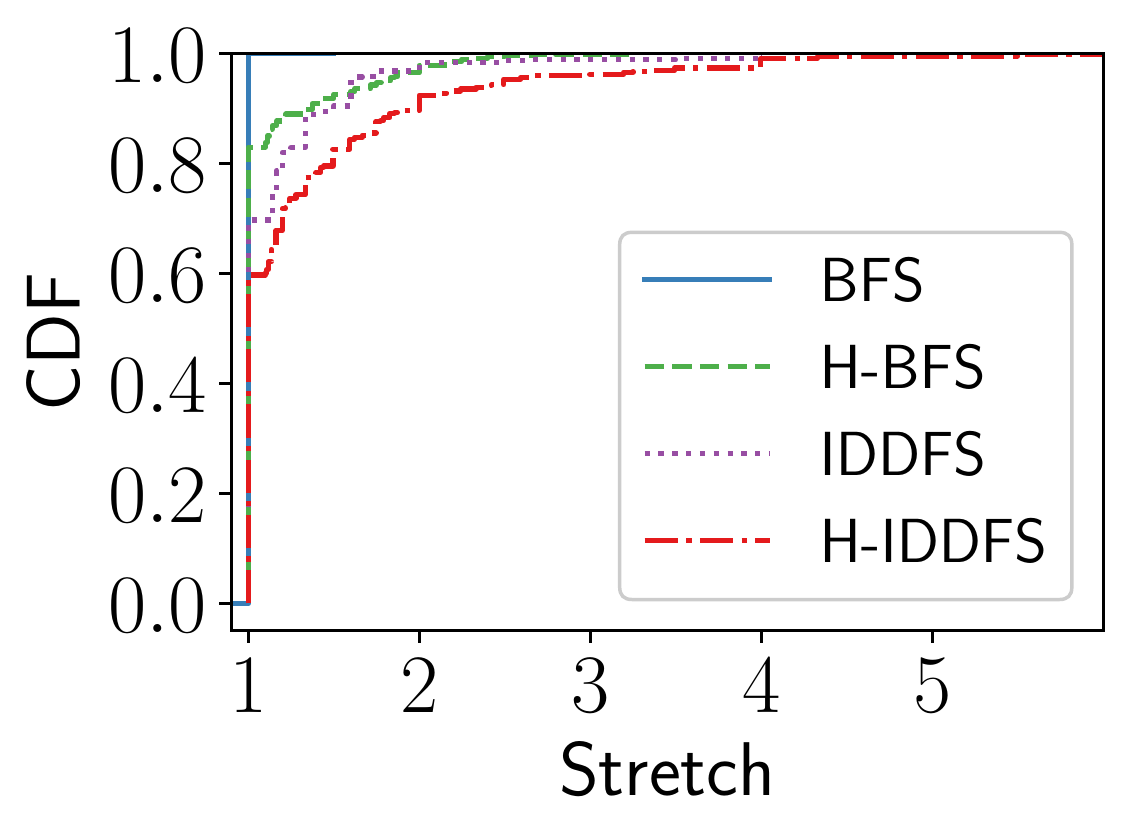}}
    \subfloat[Cesnet-Stretch]{\includegraphics[width=0.5\columnwidth]{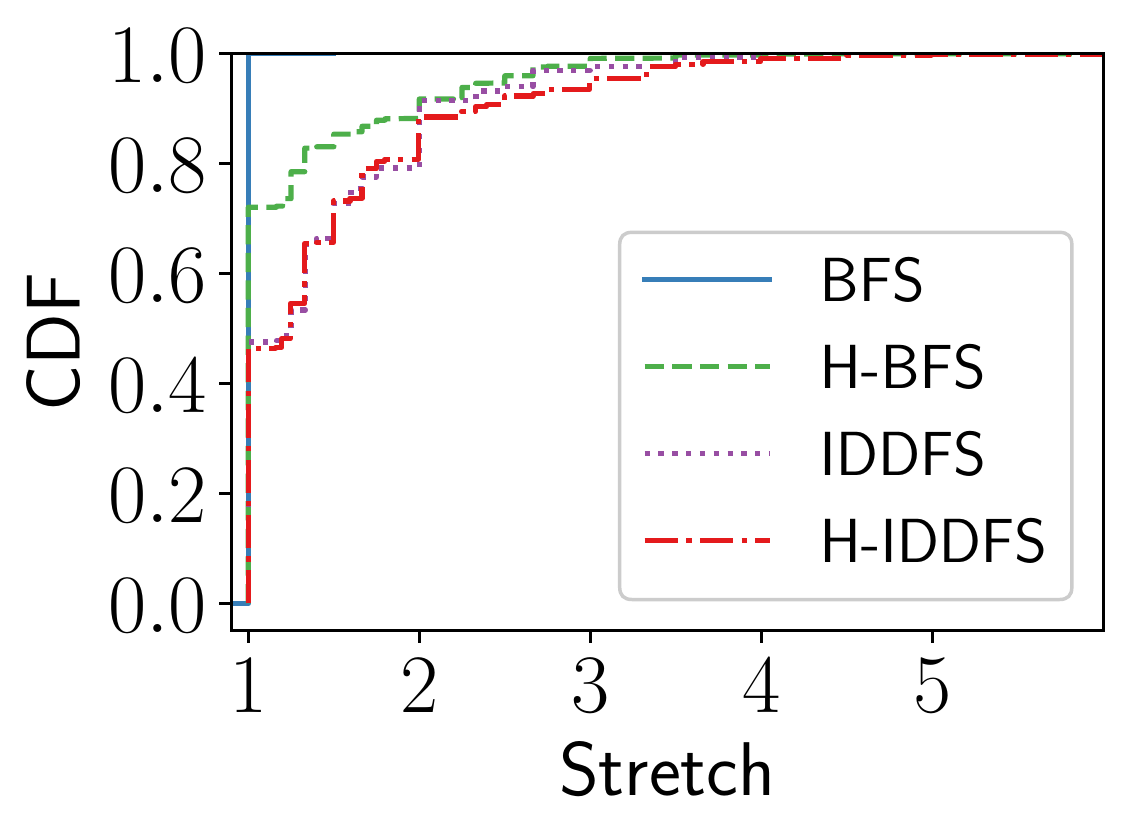}}
    \vspace*{-3mm}
    \caption{\label{fig:hierarchical} \footnotesize
        Recirculation and stretch CDF for BFS and IDDFS with and without hierarchy for two networks - 
        NetworkUsa (78 links) and Cesnet (126 links). H-BFS and H-IDDFS represent the hierarchical schemes.}
    \vspace*{-3mm}
\end{figure*}

\section{Implementation and Evaluation} \label{sec:evaluation}
The implementation of the \name data plane consists of ${\sim}$2000 lines of
P4$_{16}$ defining the P4 software behavioral model~\cite{bmv2} that emulates
the behavior of programmable switch architectures. The implementation of the
policy plane consists of ${\sim}$3000 lines of Python that use the topology
specification to generate the table rules for each P4 switch in the topology.
The policy plane hands off these rules to the switch control plane which uses
the switch APIs to install the table rules in the software bmv2 and hardware
switch. 

Both of our graph traversal algorithms (i.e., BFS and IDDFS)  use 10 stages to
run the tables outlined in \secref{sec:dataplanerouting}. Note that the number
of stages is configurable based on the switch resource requirements and other
potential applications running on the switch (e.g., firewalls, ACLs etc.). In
our experiments, we store 8 hops in the source route of the packet.

We evaluate the effectiveness of routing using \name using the Internet Zoo
topologies~\cite{zoo} (5-70 switches, 10-150 links) and use
failures scenarios varying between 1 and 3 links.
We ask the following questions:
\begin{compactitemize}
    \item Can \name find paths with low path stretch and few recirculations for 
    different topologies under different failure scenarios? (\secref{sec:zooeval})
    \item Can \name hierarchical routing reduce the number of  recirculation? 
    How do domain sizes affect the number of recirculation and the path stretch? (\secref{sec:hierarchicaleval})
    \item Can \name provide end-to-end connectivity in an emulated 
    network and on real hardware? (\secref{sec:realeval})
\end{compactitemize}
In hardware, switches generate a packet to indicate a link is
down and the delay between the actual link failure and packet generation is a
few microseconds. At 10Gbps, the data loss occurring between actual failure event
and the data plane reacting to the failure will be in the order of kilobits,
i.e, 1-2 packets (thus, nearly zero drops). For the rest of the section, we
assume that failure detection is instantaneous.

\subsection{Routing Effectiveness} \label{sec:zooeval}
In this section, we evaluate \name's ability to find routes using BFS and IDDFS,
and measure the number of recirculations and path stretch incurred by both
techniques. For these experiments, we generate packets for all pairs of
endpoints in the network and emulate the data plane behavior using bmv2. We
simulate the network by analyzing the output packet from bmv2 and "forwarding"
it to next switch. 
To evaluate \name under
failures, we generate 20 failure scenarios for each number of failed links
$k=\{1, 2, 3\}$, and observe the routing behavior for all-to-all traffic. 

Even
with source routing, packets could undergo route computations at multiple
switches (either due to failures or partially stored paths), thus, we report the
total network recirculation in \Cref{fig:recirczoo}. We define the path stretch
as the ratio of the actual path taken by the packet in the \name network
compared to the shortest active path in the network (computed by an oracle using
BFS). We report the stretch for the networks in \Cref{fig:stretchzoo}.

In the absence of failures ($k=0$), \name can find routes using few
recirculations (average {<}7) for the different networks, and we observe more
recirculations as the network size increases. This increase is expected; we need
more table invocations to explore the switches and links to find a route.
Recirculations are mostly confined to the source switch that stores the path (8
hops); the remaining switches can forward the packet to the destination based on
the source route. We observe stretch of 1 for BFS because it finds the shortest
path. For our experiments, we start IDDFS with maximum length as 4 and increase
step size by a factor of 2. Thus, IDDFS incurs a higher stretch as it does not
always find the shortest paths (though the stretch will be bounded). By
configuring the starting depth and increments, we can achieve stretch comparable
to BFS).

In the presence of failures ($k>0$), the FCP algorithm kicks in and \name needs
to recompute paths on multiple switches as packets learn about new link
failures. Moreover, a switch does not have the full view of the current topology
and it may compute a route through a failed link which the packet has not seen
yet, resulting in higher path stretch. As expected, we observe more
recirculations and higher stretch when failures occur. We also observe higher
variance in recirculation compared to $k=0$ because routing using FCP highly
depends on the topology and failure scenario. We  do not see a significant
increase in recirculations and stretch 
with increasing $k$ as the number of packets traversing through a 
route encountering all link failures will be low. 

\paragraph{Policy Routing.} We evaluate the effect of policies 
on recirculation and stretch. Note that preferences and weighted load-balancing policies
do not incur any additional recirculation or increased stretch as they simply change the 
order in which the next-hop is explored, thus, traversal will not use additional switch stages. 
We evaluate the number of recirculations and stretch 
for middlebox policies in \Cref{fig:middlebox} for different topologies under no failures. 
We generate 100 policies for random endpoints which traverse one middlebox (chosen at random), 
and the data plane computes a route from source to destination through the middlebox. 
We observe higher recirculations as a middlebox policy is effectively two traversals 
in the switch. Path stretch trends are similar to routing without policies, as the shortest 
compliant path also becomes longer. 

In summary, we show that both BFS and IDDFS can perform routing with $<10$ recirculations on average,
even under failures. Path stretch under failures is less than 2, thus, FCP does not incur 
very high stretch despite having a partial view of failure information.

\begin{figure}
	\centering
	\subfloat[\# Recirculations]{\includegraphics[width=0.47\columnwidth]{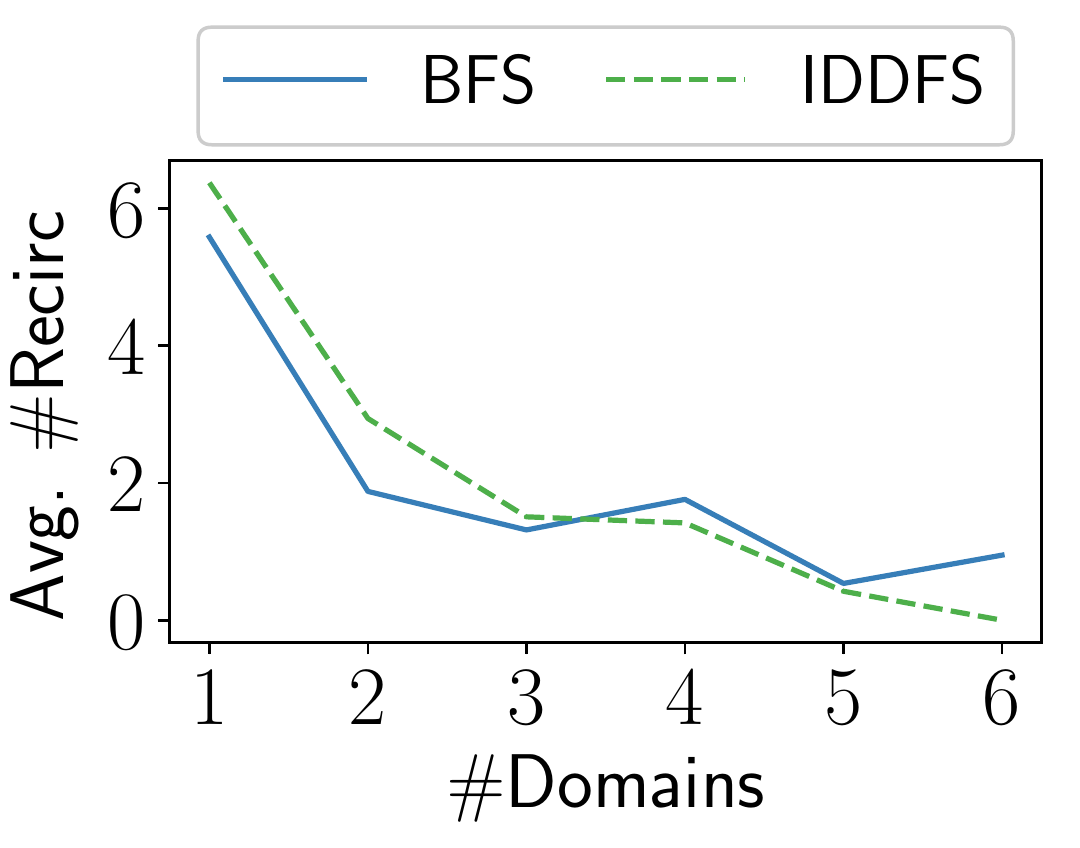}}
    \subfloat[Stretch]{\includegraphics[width=0.52\columnwidth]{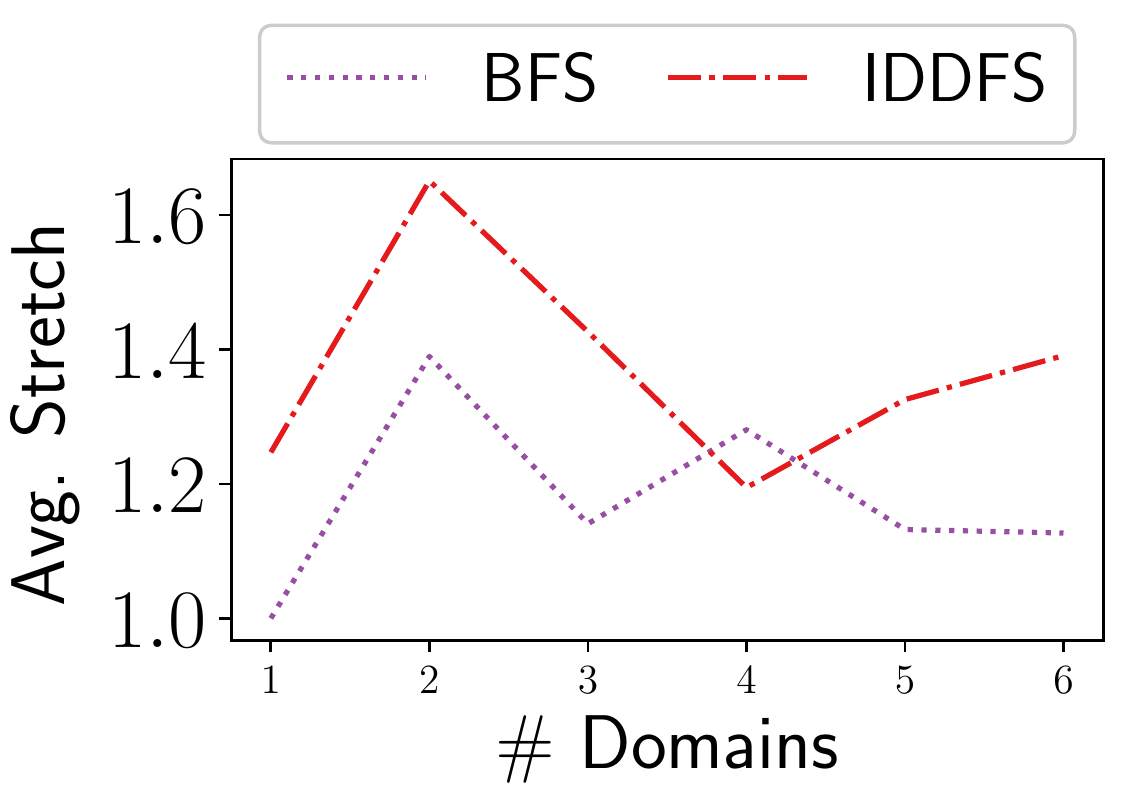}}
    \caption{Average recirculation and stretch for NetworkUSA (35 switches, 78 links) with varying 
    number of domains.}
	\label{fig:hierarchydomains}
\end{figure}

\subsection{Hierarchical Routing} \label{sec:hierarchicaleval}
We now evaluate if we can reduce recirculations using hierarchical routing in terms of
number of recirculations.
For clarity of exposition, we focus on two topologies from Internet Zoo - NetworkUsa with 35 switches and 78 links,
and Cesnet (201006) with 52 switches and 126 links. 
For both
networks, we partition the network randomly into 3 domains of 
roughly equal size. For endpoints residing in
the same domain, \name will not perform hierarchical routing. 
We consider all pairs of switches
as endpoints and plot the cumulative distribution of the total network
recirculation in \Cref{fig:hierarchical}(a,b) and path stretch in 
\Cref{fig:hierarchical}(c,d) for all routing strategies. 

Hierarchical routing results in a significant reduction in recirculation for both networks and
for both BFS and IDDFS:  the maximum recirculation suffered in Cesnet with hierarchical
routing is 1, compared to ${\geq}10$ recirculations without hierarchies. 
Remarkably, hierarchical routing
achieves 0 recirculations for the majority of endpoints in both
networks (50-60\% in NetworkUsa, 75-80\% in Cesnet). 
Finally, most of the traffic 
suffers low stretch even with hierarchical routing (80\% traffic have a 
stretch ${<}2$ with hierarchies).

We conclude this section by evaluating the effect that varying the number of
domains has on the number of recirculations and path stretch
(\Cref{fig:hierarchydomains}). As the number of domains increases, the average
recirculations decreases (more domains implies each domain is smaller, thus less
computation). The effect on stretch with changing domains is harder to analyze,
as the stretch depends on the topology structure and how the domains are
assigned. However, we note that generally higher stretch is incurred when the
number of domains is $>1$ because a switch computes routes on a
partial topology. 

In summary, hierarchical BFS and IDDFS are able to eliminate recirculations for 
majority of the endpoints without a significant increase in average 
path stretch (1.2-1.6$\times$).

\subsection{\name in the Real World}
\label{sec:realeval}
We run \name on a Stordis BF606X switch which can run P4 Tofino programs. The
first aspect of running \name on hardware is compiling to Tofino. We use
Barefoot P4 Studio~\cite{p4studio} to compile a version of \name that adheres to
the resource constraints of the switch. Since we only possess a single hardware
switch, we could not perform an end-to-end routing demonstration using \name. To
understand the viability of \name, we study the effects of recirculation on
Tofino. By configuring adequate ports in loopback for recirculation, we are able
to run \name on the switch and can perform 7 recirculations with minimal
degradation in throughput (< 1\%), and additional latency in the
order of microseconds between two hosts connected to the switch. We do not
report actual numbers due to a confidentiality agreement with Barefoot. 

\paragraph{End-to-end connectivity using Mininet.} 
We demonstrate end-to-end routing of \name using an emulated
Mininet~\cite{mininet} network of four P4 switches and two hosts
(\Cref{fig:bfsexample}). The P4 switches run the  P4 software behavioral
model~\cite{bmv2}. The bmv2 CLI is used to program the switch rules for each
switch for routing and forwarding  packets in the network. We send UDP traffic
from $S$ to $T$ with the \name headers as payload.  We disable link $1-2$ (using
a link failure status register in the switch) and bring it back up after some
time. 

Initially, the  switch 1 data plane finds a path using IDDFS ($1 \rightarrow 2
\rightarrow 4$) and the rest of the switches forward along the path stored in
the packet header. When $1-2$ has failed, 1 successfully finds an alternate path
$1\rightarrow 3\rightarrow 4$ without any packet drops (assuming failure
detection is instantaneous, emulating instantaneous failure detection). Finally,
when link $1-2$ is back up and the switch failure state is updated, packets once
again switch to path $1\rightarrow 2\rightarrow 4$. We also verify that switch 2
and 3 do not compute the paths, instead use the route installed in the packet
header.


\section{Related Work}
Blink~\cite{blink} is a state-of-art data-driven data plane solution for
connectivity recovery. Blink analyzes TCP-induced signals to detect remote link
failures that disrupt end-to-end connectivity. Once Blink has detected a remote
link failure, it uses a data-driven fast reroute mechanism: it  probes all next
hops for availability and chooses a working one. However, without any topology
information, Blink cannot fundamentally prevent forwarding issues like
blackholes. \name, on the other hand, does not actively detect remote
failures and instead uses FCP for failure propagation. \name could be
potentially used as Blink's reroute mechanism.

Sedar et. al~\cite{p4frr} propose a local fast reroute mechanism to deal with
link failures. In the data-plane FRR primitive, the packet keeps track of all
ports it has used in an attempt to reach the destination, and the data plane
sends the packet on the next available port. To re-establish connectivity, they
implement multiple mechanisms that leverage the FRR primitive to explore paths
in the network using different strategies---e.g., Rotor-Router, DFS and BFS. The
authors advocate for FRR as it does not incur recirculations (FRR is implemented
using one table) and uses less resources. We argue that \name can be used to
implement routing itself in the data plane without incurring considerable
overheads, thus, eliminating the need for local fast reroute mechanisms which can
consume network bandwidth to explore paths.

Molero et. al~\cite{hwpathvector} propose a path vector protocol using
programmable switches, and offloading key control plane functionalities to the
data plane, in the same vein as our vision. However, a distributed path vector
protocol, even one accelerated by hardware, will suffer losses during routing
convergence periods. Similarly, path vector protocol cannot easily guarantee
policy-compliance under failures and will require control plane intervention. 

Finally, one of the major avenues of research orthogonal to work is leveraging
programmable data planes to perform various in-network computing tasks
efficiently: key-value stores~\cite{netcache}, scale-free coordination for
distributed systems~\cite{netchain}, stateful load balancers~\cite{silkroad},
network ordering for consensus~\cite{nopaxos}, heavy hitter
detection~\cite{heavyhitter}, and distributed aggregation for machine
learning~\cite{p4ml}. We could potentially run \name and these applications in
parallel in the same data plane with \name performing routing while the
applications act on other packet headers. 


\section{Conclusion}
We present \name, a new network architecture that leverages programmable
switching technologies to perform routing completely in the data plane using P4. 
\name is able to 
provide always-availability and policy-compliance under failures. 
Our work opens up a vast avenue of interesting open problems: 
With current programmable switch architectures, can we implement 
shortest path algorithm for weighted graphs (to mimic
OSPF/BGP configurations)? 
Can we increase the coverage of policies we can
implement in the data plane? 
Can we design hardware optimized for graph 
traversal to perform routing efficiently?


\bibliographystyle{plain}
\bibliography{references}

\begin{thebibliography}{10}

\bibitem{bmv2}
Behavioral model repository.
\newblock \url{https://github.com/p4lang/behavioral-model}.

\bibitem{frr}
Ipv4 loop-free alternate fast reroute.
\newblock
  \url{https://www.cisco.com/c/en/us/td/docs/ios-xml/ios/iproute_pi/configuration/xe-3s/iri-xe-3s-book/iri-ip-lfa-frr.html}.

\bibitem{mininet}
Mininet.
\newblock \url{http://mininet.org/}.

\bibitem{p4studio}
P4studio.
\newblock \url{https://barefootnetworks.com/products/brief-p4-studio/}.

\bibitem{netkat}
Carolyn~Jane Anderson, Nate Foster, Arjun Guha, Jean-Baptiste Jeannin, Dexter
  Kozen, Cole Schlesinger, and David Walker.
\newblock Netkat: Semantic foundations for networks.
\newblock In {\em Proceedings of the 41st ACM SIGPLAN-SIGACT Symposium on
  Principles of Programming Languages}, POPL '14, pages 113--126, New York, NY,
  USA, 2014. ACM.

\bibitem{propane}
Ryan Beckett, Ratul Mahajan, Todd Millstein, Jitu Padhye, and David Walker.
\newblock Don't mind the gap: Bridging network-wide objectives and device-level
  configurations.
\newblock In {\em Proceedings of the ACM SIGCOMM 2016 Conference on SIGCOMM},
  SIGCOMM '16, 2016.

\bibitem{loopfree1}
Jochen Behrens and J.~J. Garcia-Luna-Aceves.
\newblock Distributed, scalable routing based on link-state vectors.
\newblock In {\em Proceedings of the Conference on Communications
  Architectures, Protocols and Applications}, SIGCOMM '94, pages 136--147, New
  York, NY, USA, 1994. ACM.

\bibitem{p4}
Pat Bosshart, Dan Daly, Glen Gibb, Martin Izzard, Nick McKeown, Jennifer
  Rexford, Cole Schlesinger, Dan Talayco, Amin Vahdat, George Varghese, et~al.
\newblock P4: Programming protocol-independent packet processors.
\newblock {\em ACM SIGCOMM Computer Communication Review}, 44(3):87--95, 2014.

\bibitem{rmt}
Pat Bosshart, Glen Gibb, Hun-Seok Kim, George Varghese, Nick McKeown, Martin
  Izzard, Fernando Mujica, and Mark Horowitz.
\newblock Forwarding metamorphosis: Fast programmable match-action processing
  in hardware for sdn.
\newblock In {\em Proceedings of the ACM SIGCOMM 2013 Conference on SIGCOMM},
  SIGCOMM '13, pages 99--110, New York, NY, USA, 2013. ACM.

\bibitem{andromeda}
Michael Dalton, David Schultz, Jacob Adriaens, Ahsan Arefin, Anshuman Gupta,
  Brian Fahs, Dima Rubinstein, Enrique~Cauich Zermeno, Erik Rubow,
  James~Alexander Docauer, et~al.
\newblock Andromeda: performance, isolation, and velocity at scale in cloud
  network virtualization.
\newblock In {\em 15th $\{$USENIX$\}$ Symposium on Networked Systems Design and
  Implementation ($\{$NSDI$\}$ 18)}, pages 373--387, 2018.

\bibitem{synet}
Ahmed El-Hassany, Petar Tsankov, Laurent Vanbever, and Martin Vechev.
\newblock Network-wide configuration synthesis.
\newblock In {\em 29th International Conference on Computer Aided Verification,
  Heidelberg, Germany, 2017}, CAV'17, 2017.

\bibitem{flowtags}
Seyed~Kaveh Fayazbakhsh, Vyas Sekar, Minlan Yu, and Jeffrey~C Mogul.
\newblock Flowtags: enforcing network-wide policies in the presence of dynamic
  middlebox actions.
\newblock In {\em Proceedings of the second ACM SIGCOMM workshop on Hot topics
  in software defined networking}, pages 19--24. ACM, 2013.

\bibitem{azurefpga}
Daniel Firestone, Andrew Putnam, Sambhrama Mundkur, Derek Chiou, Alireza
  Dabagh, Mike Andrewartha, Hari Angepat, Vivek Bhanu, Adrian Caulfield, Eric
  Chung, Harish~Kumar Chandrappa, Somesh Chaturmohta, Matt Humphrey, Jack
  Lavier, Norman Lam, Fengfen Liu, Kalin Ovtcharov, Jitu Padhye, Gautham
  Popuri, Shachar Raindel, Tejas Sapre, Mark Shaw, Gabriel Silva, Madhan
  Sivakumar, Nisheeth Srivastava, Anshuman Verma, Qasim Zuhair, Deepak Bansal,
  Doug Burger, Kushagra Vaid, David~A. Maltz, and Albert Greenberg.
\newblock Azure accelerated networking: Smartnics in the public cloud.
\newblock In {\em Proceedings of the 15th USENIX Conference on Networked
  Systems Design and Implementation}, NSDI'18, pages 51--64, Berkeley, CA, USA,
  2018. USENIX Association.

\bibitem{loopfree3}
Pierre Francois and Olivier Bonaventure.
\newblock Avoiding transient loops during igp convergence in ip networks.
\newblock In {\em Proceedings IEEE 24th Annual Joint Conference of the IEEE
  Computer and Communications Societies.}, volume~1, pages 237--247. IEEE,
  2005.

\bibitem{igpconvergence}
Pierre Francois, Clarence Filsfils, John Evans, and Olivier Bonaventure.
\newblock Achieving sub-second igp convergence in large ip networks.
\newblock {\em ACM SIGCOMM Computer Communication Review}, 35(3):35--44, 2005.

\bibitem{loopfree2}
Jose~J Garcia-Lunes-Aceves.
\newblock Loop-free routing using diffusing computations.
\newblock {\em IEEE/ACM transactions on networking}, 1(1):130--141, 1993.

\bibitem{opennf}
Aaron Gember-Jacobson, Raajay Viswanathan, Chaithan Prakash, Robert Grandl,
  Junaid Khalid, Sourav Das, and Aditya Akella.
\newblock Opennf: Enabling innovation in network function control.
\newblock In {\em ACM SIGCOMM Computer Communication Review}, volume~44, pages
  163--174. ACM, 2014.

\bibitem{datacenterfailures}
Phillipa Gill, Navendu Jain, and Nachiappan Nagappan.
\newblock Understanding network failures in data centers: Measurement,
  analysis, and implications.
\newblock In {\em Proceedings of the ACM SIGCOMM 2011 Conference}, SIGCOMM '11,
  pages 350--361, New York, NY, USA, 2011. ACM.

\bibitem{controlplanelatency}
Keqiang He, Junaid Khalid, Aaron Gember-Jacobson, Sourav Das, Chaithan Prakash,
  Aditya Akella, Li~Erran Li, and Marina Thottan.
\newblock Measuring control plane latency in sdn-enabled switches.
\newblock In {\em Proceedings of the 1st ACM SIGCOMM Symposium on Software
  Defined Networking Research}, SOSR '15, pages 25:1--25:6, New York, NY, USA,
  2015. ACM.

\bibitem{blink}
Thomas Holterbach, Edgar~Costa Molero, Maria Apostolaki, Alberto Dainotti,
  Stefano Vissicchio, and Laurent Vanbever.
\newblock Blink: Fast connectivity recovery entirely in the data plane.
\newblock In {\em 16th {USENIX} Symposium on Networked Systems Design and
  Implementation ({NSDI} 19)}, pages 161--176, Boston, MA, February 2019.
  {USENIX} Association.

\bibitem{swan}
Chi-Yao Hong, Srikanth Kandula, Ratul Mahajan, Ming Zhang, Vijay Gill, Mohan
  Nanduri, and Roger Wattenhofer.
\newblock Achieving high utilization with software-driven wan.
\newblock In {\em Proceedings of the ACM SIGCOMM 2013 Conference on SIGCOMM},
  SIGCOMM '13, pages 15--26, New York, NY, USA, 2013. ACM.

\bibitem{b4}
Sushant Jain, Alok Kumar, Subhasree Mandal, Joon Ong, Leon Poutievski, Arjun
  Singh, Subbaiah Venkata, Jim Wanderer, Junlan Zhou, Min Zhu, et~al.
\newblock B4: Experience with a globally-deployed software defined wan.
\newblock In {\em ACM SIGCOMM Computer Communication Review}, volume~43, pages
  3--14. ACM, 2013.

\bibitem{netchain}
Xin Jin, Xiaozhou Li, Haoyu Zhang, Nate Foster, Jeongkeun Lee, Robert
  Soul{\'e}, Changhoon Kim, and Ion Stoica.
\newblock Netchain: Scale-free sub-rtt coordination.
\newblock In {\em 15th {USENIX} Symposium on Networked Systems Design and
  Implementation ({NSDI} 18)}, pages 35--49, Renton, WA, April 2018. {USENIX}
  Association.

\bibitem{netcache}
Xin Jin, Xiaozhou Li, Haoyu Zhang, Robert Soul{\'e}, Jeongkeun Lee, Nate
  Foster, Changhoon Kim, and Ion Stoica.
\newblock Netcache: Balancing key-value stores with fast in-network caching.
\newblock In {\em Proceedings of the 26th Symposium on Operating Systems
  Principles}, SOSP '17, pages 121--136, New York, NY, USA, 2017. ACM.

\bibitem{compiling-dependencies}
Lavanya Jose, Lisa Yan, George Varghese, and Nick McKeown.
\newblock Compiling packet programs to reconfigurable switches.
\newblock In {\em 12th {USENIX} Symposium on Networked Systems Design and
  Implementation ({NSDI} 15)}, pages 103--115, Oakland, CA, May 2015. {USENIX}
  Association.

\bibitem{bfd}
D.~Katz and D.~Ward.
\newblock {Bidirectional Forwarding Detection (BFD)}.
\newblock {RFC} 5880.

\bibitem{zoo}
S.~Knight, H.X. Nguyen, N.~Falkner, R.~Bowden, and M.~Roughan.
\newblock The internet topology zoo.
\newblock {\em Selected Areas in Communications, IEEE Journal on}, 29(9):1765
  --1775, october 2011.

\bibitem{convergenceloss}
Craig Labovitz, Abha Ahuja, Abhijit Bose, and Farnam Jahanian.
\newblock Delayed internet routing convergence.
\newblock {\em ACM SIGCOMM Computer Communication Review}, 30(4):175--187,
  2000.

\bibitem{fcp}
Karthik Lakshminarayanan, Matthew Caesar, Murali Rangan, Tom Anderson, Scott
  Shenker, and Ion Stoica.
\newblock Achieving convergence-free routing using failure-carrying packets.
\newblock In {\em Proceedings of the 2007 Conference on Applications,
  Technologies, Architectures, and Protocols for Computer Communications},
  SIGCOMM '07, pages 241--252, New York, NY, USA, 2007. ACM.

\bibitem{nopaxos}
Jialin Li, Ellis Michael, Naveen~Kr Sharma, Adriana Szekeres, and Dan~RK Ports.
\newblock Just say $\{$NO$\}$ to paxos overhead: Replacing consensus with
  network ordering.
\newblock In {\em 12th $\{$USENIX$\}$ Symposium on Operating Systems Design and
  Implementation ($\{$OSDI$\}$ 16)}, pages 467--483, 2016.

\bibitem{ddc}
Junda Liu, Baohua Yan, Scott Shenker, and Michael Schapira.
\newblock Data-driven network connectivity.
\newblock In {\em Proceedings of the 10th ACM Workshop on Hot Topics in
  Networks}, page~8. ACM, 2011.

\bibitem{updatesynthesis}
Jedidiah McClurg, Hossein Hojjat, Pavol \v{C}ern\'{y}, and Nate Foster.
\newblock Efficient synthesis of network updates.
\newblock In {\em Proceedings of the 36th ACM SIGPLAN Conference on Programming
  Language Design and Implementation}, PLDI '15, pages 196--207, New York, NY,
  USA, 2015. ACM.

\bibitem{silkroad}
Rui Miao, Hongyi Zeng, Changhoon Kim, Jeongkeun Lee, and Minlan Yu.
\newblock Silkroad: Making stateful layer-4 load balancing fast and cheap using
  switching asics.
\newblock In {\em Proceedings of the Conference of the ACM Special Interest
  Group on Data Communication}, pages 15--28. ACM, 2017.

\bibitem{hwpathvector}
Edgar~Costa Molero, Stefano Vissicchio, and Laurent Vanbever.
\newblock Hardware-accelerated network control planes.
\newblock In {\em Proceedings of the 17th ACM Workshop on Hot Topics in
  Networks}, HotNets '18, pages 120--126, New York, NY, USA, 2018. ACM.

\bibitem{decentralizedupdate}
Thanh~Dang Nguyen, Marco Chiesa, and Marco Canini.
\newblock Decentralized consistent updates in sdn.
\newblock In {\em Proceedings of the Symposium on SDN Research}, SOSR '17,
  pages 21--33, New York, NY, USA, 2017. ACM.

\bibitem{e2}
Shoumik Palkar, Chang Lan, Sangjin Han, Keon Jang, Aurojit Panda, Sylvia
  Ratnasamy, Luigi Rizzo, and Scott Shenker.
\newblock E2: a framework for nfv applications.
\newblock In {\em Proceedings of the 25th Symposium on Operating Systems
  Principles}, pages 121--136. ACM, 2015.

\bibitem{simple}
Zafar~Ayyub Qazi, Cheng-Chun Tu, Luis Chiang, Rui Miao, Vyas Sekar, and Minlan
  Yu.
\newblock Simple-fying middlebox policy enforcement using {SDN}.
\newblock In {\em Proceedings of the ACM SIGCOMM 2013 Conference on SIGCOMM},
  SIGCOMM '13, pages 27--38, New York, NY, USA, 2013. ACM.

\bibitem{kinetic}
Mark Reitblatt, Nate Foster, Jennifer Rexford, Cole Schlesinger, and David
  Walker.
\newblock Abstractions for network update.
\newblock In {\em Proceedings of the ACM SIGCOMM 2012 Conference on
  Applications, Technologies, Architectures, and Protocols for Computer
  Communication}, SIGCOMM '12, pages 323--334, New York, NY, USA, 2012. ACM.

\bibitem{waferthincontrolplane}
Jennifer Rexford, Albert Greenberg, Gisli Hjalmtysson, David~A. Maltz, Andy
  Myers, Geoffrey Xie, Jibin Zhan, and Hui Zhang.
\newblock Network-wide decision making: Toward a wafer-thin control plane.
\newblock In {\em In HotNets-III}, 2004.

\bibitem{p4ml}
Amedeo Sapio, Ibrahim Abdelaziz, Abdulla Aldilaijan, Marco Canini, and Panos
  Kalnis.
\newblock In-network computation is a dumb idea whose time has come.
\newblock In {\em Proceedings of the Sixteenth ACM Workshop on Hot Topics in
  Networks}, 2017.

\bibitem{p4frr}
Roshan Sedar, Michael Borokhovich, Marco Chiesa, Gianni Antichi, and Stefan
  Schmid.
\newblock Supporting emerging applications with low-latency failover in p4.
\newblock In {\em Proceedings of the 2018 Workshop on Networking for Emerging
  Applications and Technologies}, NEAT '18, pages 52--57, New York, NY, USA,
  2018. ACM.

\bibitem{heavyhitter}
Vibhaalakshmi Sivaraman, Srinivas Narayana, Ori Rottenstreich, Shan
  Muthukrishnan, and Jennifer Rexford.
\newblock Heavy-hitter detection entirely in the data plane.
\newblock In {\em Proceedings of the Symposium on SDN Research}, pages
  164--176. ACM, 2017.

\bibitem{merlin}
Robert Soul{\'e}, Shrutarshi Basu, Parisa~Jalili Marandi, Fernando Pedone,
  Robert Kleinberg, Emin~Gun Sirer, and Nate Foster.
\newblock Merlin: A language for provisioning network resources.
\newblock In {\em Proceedings of the 10th ACM International on Conference on
  Emerging Networking Experiments and Technologies}, CoNEXT '14, pages
  213--226, New York, NY, USA, 2014. ACM.

\bibitem{genesis}
Kausik Subramanian, Loris D'Antoni, and Aditya Akella.
\newblock Genesis: Synthesizing forwarding tables for multi-tenant networks.
\newblock In {\em POPL}. ACM, 2017.

\bibitem{zeppelin}
Kausik Subramanian, Loris D'Antoni, and Aditya Akella.
\newblock Synthesis of fault-tolerant distributed router configurations.
\newblock {\em Proceedings of the ACM on Measurement and Analysis of Computing
  Systems}, 2(1):22, 2018.

\bibitem{fibbing}
Stefano Vissicchio, Olivier Tilmans, Laurent Vanbever, and Jennifer Rexford.
\newblock Central control over distributed routing.
\newblock In {\em Proceedings of the 2015 ACM Conference on Special Interest
  Group on Data Communication}, SIGCOMM '15, pages 43--56, New York, NY, USA,
  2015. ACM.

\bibitem{wcmp}
Junlan Zhou, Malveeka Tewari, Min Zhu, Abdul Kabbani, Leon Poutievski, Arjun
  Singh, and Amin Vahdat.
\newblock Wcmp: Weighted cost multipathing for improved fairness in data
  centers.
\newblock In {\em Proceedings of the Ninth European Conference on Computer
  Systems}, page~5. ACM, 2014.

\bibitem{timegone}
Noa Zilberman, Matthew Grosvenor, Diana~Andreea Popescu, Neelakandan
  Manihatty-Bojan, Gianni Antichi, Marcin W{\'o}jcik, and Andrew~W Moore.
\newblock Where has my time gone?
\newblock In {\em International Conference on Passive and Active Network
  Measurement}, pages 201--214. Springer, 2017.

\end{thebibliography}

\appendix

\section{Failure Carrying Packets} \label{sec:fcp}
Failure Carrying Packets (FCP)~\cite{fcp} is a distributed routing
paradigm designed to \emph{eliminate} convergence periods altogether---a packet
is guaranteed to reach the destination if a path to the destination exists in
the network. FCP takes advantage of the fact that permanent network topology
change (in terms of provisioning/de-provisioning links and switches) happens at
the timescales of weeks/months and is well-planned. The only changes for which
operators are not prepared for are links and routers failing and coming back up
at smaller timescales~\cite{datacenterfailures}. Thus, each router has a consistent
topology description which indicates all switches and adjacencies between them. 

\begin{figure}
	\centering
	\includegraphics[width=0.6\columnwidth]{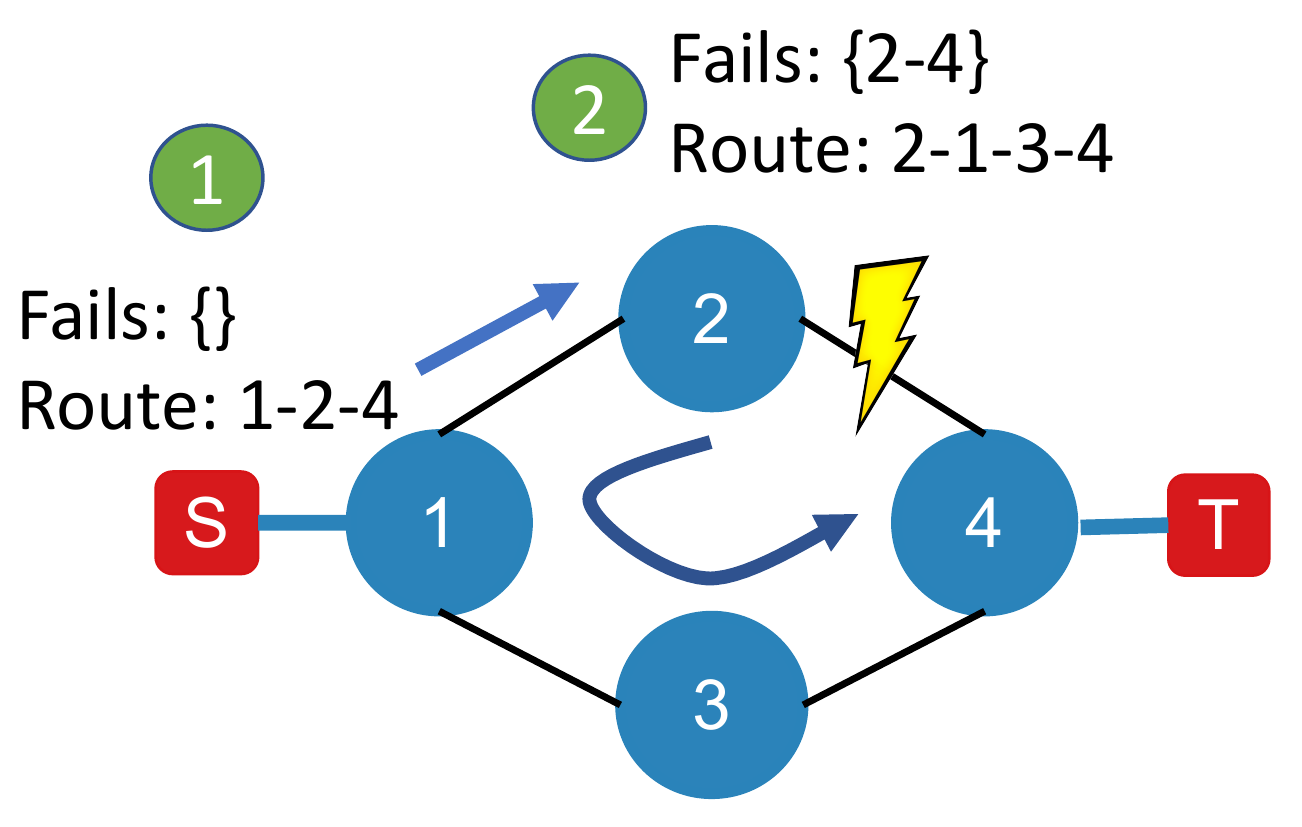}
	\caption{Example of FCP protocol in action when link 2-4 is down.}
	\label{fig:fcpexample}
\end{figure}

\begin{figure}
	\begin{minipage}{\linewidth}
		\begin{algorithm}[H]
				\caption{Failure Carrying Packet Protocol}
				\label{alg:fcp}
				\begin{algorithmic}[1]
                    \Procedure{FCP}{dst, router}
                    \State{pkt.failed_links $\bigcup$= router.failed_links}
                    \State{path = ComputePath(topo - pkt.failed_links)}
                    \If{path == null}
                        \State{// No path to destination}
                        \State{router.drop();}
                    \Else
                        \State{router.forward(pkt, path.next_hop)}
                    \EndIf
					\EndProcedure
				\end{algorithmic}
		\end{algorithm}
	\end{minipage}
\end{figure}

The intuition behind FCP is that if the switch knows the list of failed links in
the network, it can successfully route a packet to the destination using the
network topology and failure information. However, knowledge of all link
failures will require a link-state advertisement protocol, which can lead to
convergence issues. Instead, in FCP, each packet header carries information
about all failed links it has encountered, and the switch simply uses the
topology and failure information to route the packet to the destination. The
packet on the route may again encouter a failed link to the next-hop, in that
case, the failed links is added to the packet header and route is once again
recomputed at the new router, and so on. We illustrate an example of the FCP
protocol in \Cref{fig:fcpexample}. Switch 1 computes the route $1\rightarrow 2
\rightarrow 4$ (\greenc{1}) to the destination as it does not have any
information about the failed $2-4$ link. When the packet reaches switch 2, the
failure information in the packet is updated and switch 2 computes the new route
to the destination $2\rightarrow 1 \rightarrow 3 \rightarrow 4$ (\greenc{2}).
Switch 1 receives the packet once again, but it will not send the packet to 2 as
the switch knows that $2-4$ is failed, and thus, sends it to 3 and so on.

FCP is able to guarantee reachability
if a path exists by the following intuition: at every switch in the network, the
packet will monotonically increase the set of failed links in the
packet\footnote{FCP does not consider link flapping---i.e., the packet
encountered a failure and updated its header, but the link came back up before
the packet reached the destination.}. Thus eventually, the packet would get
information about all failed links in the network, and any router would be able
to route the packet to the destination if a path exists.
The only failure state maintained by an FCP router is the failure state of links
connected to the router. FCP 
learns about the state of remote links solely from the packet headers, and 
importantly, it does not store this information. Thus, FCP routers do not 
need to advertise failures unlike OSPF. Thus, while FCP can incur additional 
stretch, we can avoid the link-state flooding overhead during failures.


With programmable switch architectures, realizing a FCP-like protocol is more
practical than when FCP was actually introduced. One of the major deployment 
challenges for FCP was changing the router hardware to support 
a new protocol header to incorporate information 
about link failures. With P4, we can easily define our custom protocol header 
and parsers, which can be efficiently run on hardware at line rates. We store 
the failure information in the header as a bit-vector where each bit represents
the state of a particular link in the topology.

\section{DFS P4 Implementation} \label{sec:dfsp4}
Let's consider a switch $m$ at len $l_1$. IDDFS will explore 
a neighbor $n$  if $m \rightarrow n$ is not visited/failed and 
$l_1 < $ \texttt{max_len}. If $n$ is valid, IDDFS pushs
$m$ (\texttt{curr}) into the stack for backtracking purposes. IDDFS
will also mark the incoming edges to $n$ as visited and update 
the current path length (lines~\ref{alg:nexthop}-\ref{alg:nexthopend}). We 
add the following table rule(s) for $m$ to implement the above logic:
\begin{Verbatim}[frame=single]
match:
curr:m=1,visited_vec:*******0,len:0,max_len:4
                    ...
curr:m=1,visited_vec:*******0,len:3,max_len:4

action goto_neighbor(n, n_visited) {
  Stack.push(hdr.curr);
  hdr.curr = n;
  hdr.visited_vec = hdr.visited_vec|n_visited;
  hdr.len++;}
\end{Verbatim}
The match condition for visited_vec and the action parameters 
(\texttt{n, n_visited}) are same for BFS and IDDFS (shown in 
\Cref{fig:bfsexample}).

At any switch $m$, we backtrack (line~\ref{alg:btvisited}, ~\ref{alg:btdepth})
when either all outgoing edges of $m$ are visited/failed, or the current path
length exceeds the max length. We set backtrack as the default action of our
IDDFS table, and will be executed whenever the current header values do not
match any valid match condition corresponding to the neighbors.
\begin{Verbatim}[frame=single]
default action backtrack() {
  hdr.curr = Stack.pop();
  hdr.len--;} 
\end{Verbatim}

Finally, once we have explored all switches at \texttt{max_len} distance, 
IDDFS increases the max length by a factor of 2 and resets \texttt{curr} and 
\texttt{visited_vec} to the initial state. To match to this component
of the algorithm(lines~\ref{alg:resetstart}-\ref{alg:resetend}), we 
check for a special bottom of stack switch (0):
\begin{Verbatim}[frame=single]
match:
curr:0,visited_vec:********,len:-1,max_len:4
    
action increase_length() {
  hdr.max_len = hdr.max_len << 1; // *2
  hdr.curr = curr_init;
  hdr.visited_vec = visited_init;
  hdr.len = 0;}
\end{Verbatim}
\section{Source Routing Implementation} \label{sec:paths}
While each \name data plane is capable of computing a route to the destination,
recomputing the path at each switch will incur additional recirculations. To
prevent unnecessary recomputations, we augment our graph traversal algorithms to
store the computed route in the packet. Downstream switches can use the path in
the header and forward to the next-hop without any recomputation, except in the
scenario that the next-hop in the packet is not reachable (due to a link
failure). If the next-hop is not reachable, the switch will compute a new route
and store it in the packet header. In \name, we keep track of 8 hops in the
packet header (configurable parameter), and the last switch in the path can
recompute the path to the destination and load it into the header. 

\paragraph{BFS-SR.}
The BFS algorithm explores multiple paths from the source till it 
finds the destination. Thus, when we add each switch in the stack, 
we also need to add the current path computed to the switch into the 
stack. To do this, we define our BFS stack in P4 as follows:
\begin{Verbatim}[frame=single]
header bfs_stack_entry {
  bit<8> switch;
  bit<64> path; // store max 8 hops
}
bfs_stack_entry Stack[10];
\end{Verbatim}
We keep track of the current path and 
length in \texttt{hdr.path} and \texttt{hdr.len}
respectively 
and update the \texttt{bfs} actions to keep track of the
current computed path. 
\begin{Verbatim}[frame=single]
table bfs {
key = {...
    hdr.len: exact;}

action push_neighbor(n_2) {
  newPath = hdr.path;
  newPath[hdr.len + 1] = n_2;
  Stack[~hdr.stack].push(n_2, newPath);
  ...
}
\end{Verbatim}
While popping elements from the stack in actions \texttt{pop_stack}
and \texttt{change_stack}, we update \texttt{curr} and \texttt{path}
from the top of the BFS stack. Finally, once we have reached the 
destination in the algorithm, \texttt{hdr.path} will reflect 
the path from source to destination and is emitted in the 
deparser for use by downstream switches. 

\paragraph{IDDFS-SR.}
The IDDFS algorithm explores along a single path, backtracking till the
algorithm reaches the destination. Thus, we can store the current explored
path in \texttt{hdr.path} and do not need to store paths in the stack like 
BFS. Since, we already track the current path length, we need to modify 
the \texttt{iddfs} actions to store the path in the packet header, 
which can be then transmitted. 
\begin{Verbatim}[frame=single]
action goto_neighbor(n_2) {
  hdr.curr = n_2;
  hdr.len++;
  hdr.path[hdr.len] = n_2;}

action backtrack() {
  hdr.curr = Stack.pop();
  hdr.path[hdr.len] = 0; //erase len index
  hdr.len--;}
  
action increase_length()
  ...
  hdr.curr = curr_init;
  hdr.path = 0; // erasing all
  hdr.len = 0;}
\end{Verbatim}
Source-routing is necessary for IDDFS for correctness of routing, while BFS 
can operate without storing paths. This is because IDDFS does not always 
find the shortest path (in terms of next-hops).

\end{document}